\documentclass[a4paper,10pt]{article}

\usepackage{epsfig}
\usepackage[dvips]{color}
\usepackage{graphicx}
\usepackage{amssymb}
\usepackage{fullpage}	

\newcommand{\eref}[1]{(\ref{eq:#1})}
\newcommand{\fref}[1]{fig.~\ref{fig:#1}}
\newcommand{\Fref}[1]{Fig.~\ref{fig:#1}}
\newcommand{\sref}[1]{\S\ref{sec:#1}}
\newcommand{\tref}[1]{table~\ref{tab:#1}}

\renewcommand{\d}{\mathrm{d}}
\renewcommand{\L}{\ensuremath{\mathcal{L}}}
\renewcommand{\O}[1]{\mathcal{O}\!\left[#1\right]}

\newcommand{\hlinevspace}{\\[-10pt]}

\newcommand{\etal}{{\em et al.}}

\title{Turbulent pair separation due to multi scale stagnation point structure and its time asymmetry in two-dimensional turbulence}
\author{T.~Faber~~~~~J.~C.~Vassilicos \\
{\small Department of Aeronautics and Institute for Mathematical Sciences} \\
{\small Imperial College London, Exhibition Road, London SW7 2PG, United Kingdom}}

\begin{document}

\maketitle

\begin{abstract}
The pair separation model of Goto and Vassilicos (S Goto and J C Vassilicos, 2004, New J.Phys., 6, p.65) is revisited and placed on a sound mathematical foundation. A DNS of two dimensional homogeneous isotropic turbulence with an inverse energy cascade and a $k^{-5/3}$ power law is used to investigate properties of pair separation in two dimensional turbulence. A special focus lies on the time asymmetry observed between forward and backward separation. Application of the present model to this data suffers from finite inertial range effects and thus, conditional averaging on scales rather than on time has been employed to obtain values for the Richardson constants and their ratio. The Richardson constants for the forward and backward case are found to be $(1.066\pm 0.020)$ and $(0.999\pm 0.007)$ respectively. The ratio of Richardson constants for the backwards and forwards case is therefore $g_b/g_f = (0.92 \pm 0.03)$, and hence exhibits a qualitatively different behavior from pair separation in three dimensional turbulence, where $g_b > g_f$ (J Berg \etal~, 2006, Phys.Rev.E, 74(1), p.016304). This indicates that previously proposed explanations for this time asymmetry based on the strain tensor eigenvalues are not sufficient to describe this phenomenon in two dimensional turbulence. We suggest an alternative qualitative explanation based on the time asymmetry related to the inverse versus forward energy cascade. In two dimensional turbulence, this asymmetry manifests itself in merging eddies due to the inverse cascade, leading to the observed ratio of Richardson constants.
\end{abstract}


\section{Introduction}

Lagrangian pair separation statistics are the vital ingredient necessary to calculate concentration fluctuations and concentration covariances which is important for studying pollution dispersal, combustion processes and the spread of contaminants in a liquid or gas \cite{Durbin:1980,Fung:1998}.

One of the fundamental quantities in turbulent particle dispersion is the mean square separation of an ensemble of particle pairs $\left\langle \Delta^2(t) \right\rangle$ \cite{Fung:1998}, which will be studied in this article.

Conversely, turbulent mixing, which is the convergence of particles that were further apart at an earlier time due to a turbulent flow, can be regarded as being equivalent to time inverted dispersion. Therefore, the underlying quantity of mixing processes is the so called \textit{backwards} dispersion, that is the distribution of particle separations $\left\langle \Delta^2(t) \right\rangle$ at a time $t$ for a prescribed separation $\Delta_0$ at a later time $t_0 > t$ \cite{Corrsin:1952,Flohr:2000,Frisch:1999,Sawford:2005,Thomson:2003}.

Let us denote the forward mean square separation at time\footnote{The time refers to the lapsed time of the underlying flow of the separation process. $t=0$ denotes the time when the initial / finial separation $\Delta_0$ is fixed.} $t$ with an initial separation $\Delta_0$ at $t=0$ by $\left\langle \Delta^2(t) | \Delta_0 \right\rangle_\mathrm{fwd}$. Equivalently, the backward mean square separation shall be denoted by $\left\langle \Delta^2(-t) | \Delta_0 \right\rangle_\mathrm{bwd}$. Note that the use of $-t$ means that the time argument is actually positive, since for backward processes $t<0$. For simplicity of notation in the backwards case, the minus sign shall be dropped henceforth and it is understood that time in the backwards case always refers to times smaller than the initial flow time $t=0$. For flows whose dynamics are time-reversible, like Gaussian flows or kinematic simulations (KS), one could expect that the mean square separation of the forward and backward case  coincide and indeed, this has been shown for Gaussian flows and KS \cite{Flohr:2000,Frisch:1999,Sawford:2005}:
\begin{equation}
 \left\langle \Delta^2(t) | \Delta_0 \right\rangle_\mathrm{fwd} = \left\langle \Delta^2(t) | \Delta_0 \right\rangle_\mathrm{bwd} \, .
\end{equation}
For other flows, like e.g. turbulent flows governed by the Navier-Stokes equation, this equality cannot be assumed and indeed, it has been shown using Lagrangian stochastic models \cite{Luethi:2007,Sawford:2005}, experiment and DNS \cite{Berg:2006} that in three dimensional turbulence, backwards dispersion happens at a faster rate than forward dispersion.

In this article we extend and give a sound mathematical foundation for the pair separation model as introduced by Goto \& Vassilicos 2004 \cite[shorthand GV04]{GV04}. We start with a short summary of the GV04 model and then go on to refine and extend this model and explore its features and similarities to Richardson's distance neighbor function \cite{Richardson:1926}. Finally, we investigate the asymmetry between ``forward'' and ``backward'' pair separation in a DNS of two dimensional homogeneous isotropic turbulence, and compare the findings to those in three dimensional turbulence.

\section{Pair separation model}
\label{sec:pairsep_origmodel}

The notion of pair separation as a process of burst-like separation events has been discussed and related to the streamline topology by Goto and Vassilicos \cite{GV04} using a 2D DNS with an inverse $5/3$ energy cascade. We shall remind the reader of the basic concept and introduce some refinements.

In two dimensional multiscale flows, the basic idea is that each particle in a fluid is on a so called ``patron'' eddy. These patron eddies are coherent structures which persist for a time that is long enough to influence the dynamics of the fluid. They exist at every scale present in the flow and can be visualized by applying a coarse graining filter on the velocity field with an associated cut-off scale $\eta_c$. Each such eddy has typically an elliptic zero-acceleration point at its center. It is so called because the velocity gradient tensor $\partial \vec{u} / \partial \vec{x}$ takes a vortical form around such a zero-acceleration point when one moves in the frame where the zero-acceleration point is at rest. Between the patron eddies, one finds hyperbolic zero-acceleration points which are locally surrounded by a straining velocity gradient tensor. Both elliptic and hyperbolic zero-acceleration points are assigned an associated scale $\eta_c$ corresponding to the filtering scale. In the following, this scale $\eta_c$ will be referred to as the ``size'' of the vortex or zero-acceleration point.

The GV04 model portrays the pair separation process as a series of ``bursts'' to larger scales: Two distinct particles in a fluid share at least one patron eddy at each time (e.g. they always belong to the patron with the associated scale of $\eta_c=L_0$ where $L_0$ is the size of the system / box size / etc.). Now consider the common patron eddy with the smallest associated scale $\eta_c=\Delta$ of a pair of particles at a particular time $t$. As long as both particles belong to this patron, their separation will also be typically of the scale $\Delta$. It is now possible that at some time $t + T_\xi(\Delta)$, one of the particles encounters a hyperbolic zero-acceleration point of scale $\Delta$ which removes it from the smallest common patron, where $T_\xi(\cdot)$ is some function depending on the parameter $\xi$. After such an encounter, the smallest shared patron of the particle pair will be of size $\xi \Delta$ where $\xi > 1$. Hence, the typical pair separation will also have increased.

While this picture strictly holds when locally moving with the zero-acceleration points and in two dimensions, it is possible to extend this notion to a more general context. Given sufficient persistence of the streamlines, GV04 argue that the presented arguments hold in a global reference frame when considering the elliptic and hyperbolic {\em velocity} stagnation points (i.e. points where the fluid velocity $\vec{u}=0$) instead of the zero-acceleration points in the local frame(s). This picture is not as intuitive as the patron notion in the local frames, however it seems like the natural generalization to a global frame. Hence, when considering sufficiently persistent multi-scale flows, also in three dimensions, one should be able to assume that particles belong to structures (previously called patrons) of a certain scale $\eta_c$.

In three dimensions, these structures will not be flat eddies (like the patrons in two dimensions) but rather some kind of elongated eddies such as vorticity tubes or a patch of straining region in the velocity field. Despite not knowing the exact shape and properties of such persistent structures, they have frequently been observed in turbulence experiments and DNS \cite{Jimenez:1993,She:1991}. Therefore, one can assume that they are present and have a typical scale $\eta_c$ which does not change much during the lifetime of such a persistent structure.

\subsection{Refined model}

In the present work, we shall add to the model the possibility of two particles {\em converging} to a smaller characteristic separation $\Delta$. While, as it turns out, there seems to be no practical benefit from this addition, it strengthens the derivation and mathematical foundation of the concept.

\subsubsection{Notation and basic ideas}

Let $\Delta_n \equiv \xi^n \Delta_0$ be the separation of a particle pair with initial separation $\Delta_0$ after a succession of separating and converging burst events, where $n$ is an integer. Such a burst event can occur when at least one of the pair particles encounters a straining (hyperbolic) stagnation point. Note that $\xi$ is the characteristic ratio of successive separations following a burst event, and as such is a constant that represents the respective ratio for burst events of all individual pairs.

The probability $b_n$ of encountering a straining stagnation point of size $\Delta_n$ must be inversely proportional to the characteristic time $T_\xi(\Delta_n)$ between burst events for pairs with a separation $\approx \Delta_n$, as introduced by GV04 \cite{GV04}, and therefore proportional to the mean distance between hyperbolic stagnation points $\ell(\Delta_n) = n_s(\Delta_n)^{-1/d}$:
\begin{equation}
 b_n \propto T_\xi(\Delta_n)^{-1} \qquad \Leftrightarrow \qquad b_n \propto u'\, n_s(\Delta_n)^{1/d} \, ,
\end{equation}
where the stagnation point density $n_s(\Delta_n)$ is given by \cite{Davila:2003}
\begin{equation}
\label{eq:pairsep_stag_pt_dens}
 n_s(\eta_c) = C_s\, \L^{-d} \, \left(\frac{\L}{\eta_c}\right)^{D_s} \, ,
\end{equation}
and $\L$ is the largest scale in a multi-scale flow (typically the integral scale), $\eta_c$ is the coarse-graining scale and cannot be taken smaller than the smallest scale $\eta$ of the flow below which the power-law energy scaling fails (for this section's purposes, $\eta_c = \Delta_n$), $d$ the dimension of the flow, $D_s$ the fractal dimension of the stagnation point distribution in space and $C_s$ is the \textit{stagnation point number} which is proportional to the number of $\L$-sized stagnation points per unit volume. We assume a power spectrum $E(k) \propto k^{-\rho}$, $1<\rho<2$ for \eref{pairsep_stag_pt_dens} to be valid \cite{Davila:2003}, in which case $D_s$ has been shown to obey the relation \cite{Davila:2003}
\begin{equation}
 D_s = \frac{d\, (3-\rho)}{2} \, .
\end{equation}
For the purposes of this paper, we assume $\rho = 5/3$. Due to the dependence on the stagnation point structure, one finds that the probability of encounter, $b_n$, 
mainly depends on the exponent of the energy spectrum $\rho$:
\begin{equation}
\label{eq:def_bn}
 b_n = C_B \, C_s^{1/d} \, \frac{u'}{\L^{(\rho-1)/2}} \, \Delta_n^{(\rho-3)/2} = B \, \Delta_n^{-2/3}
\end{equation}
for $\rho=5/3$, and 
\begin{equation}
\label{eq:pairsep_B_const}
 B=C_B\,C_s^{1/d}\,u'/\L^{1/3} \, . 
\end{equation}
$C_B$ is a proportionality constant for the probability of a particle pair to encounter a stagnation point.

When a particle pair with separation $\Delta_n$ encounters a straining stagnation point of size $\Delta_n$, it is assumed to separate further with a probability $p < 1$, resulting in a separation of scale $\Delta_{n+1}$ (separating burst). Alternatively, the pair can remain at the same scale of separation with probability $1-p$. This is the burst process described in GV04 \cite{GV04}, although the probability $p$ was effectively absorbed into the coefficient $B$ of \eref{def_bn} and not discussed by them.

The converging burst process can be initiated when a particle pair with separation $\Delta_n$ encounters a straining stagnation point which is {\em of smaller scale} than the separation of the pair, i.e. $\Delta_{n-1}$. As mentioned previously in \sref{pairsep_origmodel}, both straining and elliptical stagnation points have an associated coarse-graining scale $\eta_c$. Between the eddies or structures of scale $\eta_c=\Delta_n$, and their elliptical stagnation points lie hyperbolic straining stagnation points of the same scale, which can act as a ``gateway'' into or out of a coherent structure of equal scale as shown in \fref{pairsep_SSP_convergence}. Hence, a particle pair of typical separation $\Delta_n$ needs to encounter a straining stagnation point of scale $\Delta_{n-1}$ to experience the converging burst process bringing them together to a separation of $\Delta_{n-1}$.

\begin{figure}[h!tb]
\centering
\includegraphics{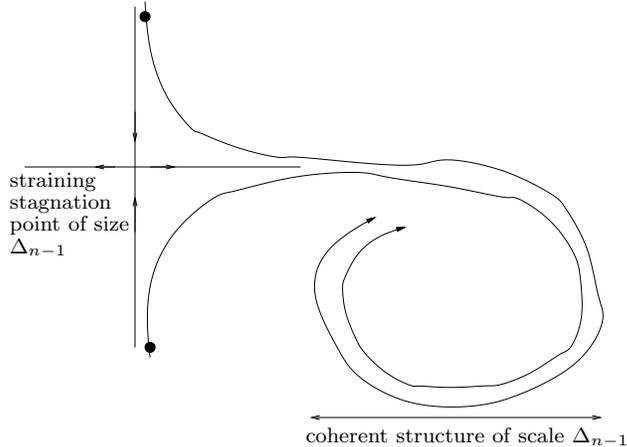}
\caption{\label{fig:pairsep_SSP_convergence}
Diagram illustrating the pair convergence process. Shown is a particle pair and their trajectories. Their separation before encountering the stagnation point is $\Delta_n$. Reversing all arrows gives the figure for the separation process. Note that the scale of the stagnation point and of the associated coherent structure is always the same, no matter whether one considers the converging or separating process.}
\end{figure}

Hence, a particle pair with separation $\Delta_n$ encounters a straining stagnation point of scale $\Delta_{n-1}$ with probability $b_{n-1}$, and then can converge to a separation of scale $\Delta_{n-1}$ with a probability called $q$.

\subsubsection{Derivation of PDF of pair separation}

Analogously to GV04 \cite{GV04}, the PDF of pair separation can be derived as follows.

Let $Q_n(t)$ be the probability for a particle pair to have a separation between $\Delta_n$ and $\Delta_{n+1}$ at a certain time $t$ and $n$ is any integer. $Q_n$ will then evolve according to the probabilities $b_n$, $b_{n-1}$, $p$ and $q$ as given by the following evolution equation,
\begin{equation}
\label{eq:pairsep_Qn_evolution}
 \frac{\d}{\d t} Q_n =  p\, b_{n-1}\, Q_{n-1} - ( p\, b_n + q\, b_{n-1} )\, Q_n + q\, b_n\, Q_{n+1} \, ,
\end{equation}
which is illustrated by \fref{pairsep_Delta}. Here, we have made use of the locality-in-scale hypothesis \cite{Fung:1998,GV04} which suggests that an increase or decrease of $Q_n$ can only originate from the neighboring separation intervals associated with the probabilties $Q_{n\pm 1}$, but not from separation intervals that are further away, like e.g. the ones associated with $Q_{n\pm i}$ with $i\geq 2$.

\begin{figure}[h!tb]
\centering
\includegraphics{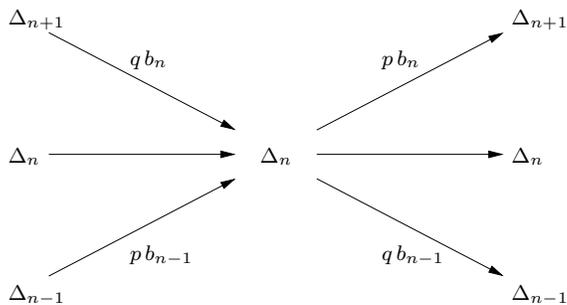}
\caption{\label{fig:pairsep_Delta}
Diagram explaining the burstwise pair separation and convergence process. Note that if the arrows were reversed (``time-reversal''), the only resulting difference would be that $p$ and $q$ are interchanged. If an encounter with a hyperbolic stagnation point does not result in a burst or convergence event, the separation of a pair remains unchanged.}
\end{figure}

In the limit of continuous separation $\Delta$, we find that
\begin{equation}
 \Delta_n \rightarrow \Delta(n) = \xi^n\, \Delta_0 \, .
\end{equation}
Defining $\alpha \equiv \ln\xi$, we arrive at
\begin{equation}
\label{eq:pairsep_Qn_P}
 Q_n = \frac{\d\Delta}{\d n}\, P(\Delta_n) = \alpha\, \Delta\, P(\Delta) \, ,
\end{equation}
where $P(\Delta)$ is the probability density function of continuous separation $\Delta$.

If one assumes that $\alpha$ is a small parameter it is possible to Taylor expand all $Q_n$ and $b_n$ in \eref{pairsep_Qn_evolution}, which results up to second order in $\alpha$ in
\begin{equation}
\label{eq:pairsep_PDF_PDE_0}
 \frac{1}{B} \: \frac{\partial P}{\partial t} = \left[ - \alpha\, (p-q) + \frac{2}{3}\, \alpha^2\, q \right] \frac{\partial}{\partial\Delta} \left(\Delta^{1/3} P\right) + \alpha^2\, \frac{p+q}{2} \, \frac{\partial}{\partial\Delta} \left(\Delta \frac{\partial}{\partial\Delta} \left(\Delta^{1/3} P\right) \right) .
\end{equation}
The index $n$ has been dropped for legibility. This result reverts to the equation found by GV04 \cite{GV04} when $q=0$ and $p=1$. This partial differential equation has 4 parameters, $B$, $p$, $q$ and $\alpha$. However, either $p$ or $q$ can be absorbed into $B$, leaving only 3 free parameters. In fact, the equation takes a tidier form if the rescaled time $\tau$ is introduced,
\begin{equation}
\label{eq:pairsep_tau_t}
 \tau = \frac{2}{9}\, B\, \alpha^2\, (p+q) \, t \, ,
\end{equation}
along with the renormalized probabilities
\begin{equation}
 p^* = \frac{p}{p+q} \qquad \mathrm{and} \qquad q^* = \frac{q}{p+q} \, .
\end{equation}
Therefore the equation now reads
\begin{equation}
\label{eq:pairsep_PDF_PDE_tidy}
 \frac{\partial P}{\partial \tau} = \left[ - \frac{9}{2} \, \frac{p^*-q^*}{\alpha} + 3\, q^* \right] \frac{\partial}{\partial\Delta} \left(\Delta^{1/3} P\right) + \frac{9}{4} \, \frac{\partial}{\partial\Delta} \left(\Delta \frac{\partial}{\partial\Delta} \left(\Delta^{1/3} P\right) \right) \, ,
\end{equation}
under the constraint $p^*+q^*=1$, leaving effectively only two free parameters (e.g. $q^*$ and $\alpha$). The third free parameter has been absorbed into the time variable, and is therefore implicitly still present.

This form highlights the influence of the model parameters on the shape of the PDF: The two free parameters occur as a combination in the PDF, hence it will not be possible to determine both parameters from the shape of the PDF. This becomes even more obvious when the equation is written in the following form:
\begin{equation}
\label{eq:pairsep_PDF_PDE_ks}
 \frac{\partial P}{\partial \tau} = k_0\,\Delta^{-2/3}\, P + k_1\,\Delta^{1/3}\, \frac{\partial P}{\partial\Delta} + k_2\,\Delta^{4/3}\, \frac{\partial^2 P}{\partial\Delta^2} \, ,
\end{equation}
where
\begin{eqnarray}
\label{eq:pairsep_k0_pq}
 k_0 &=& -\frac{3}{2}\,\frac{p^*-q^*}{\alpha} + q^* + \frac{1}{4} \, , \\
 k_1 &=& 3\, (k_0+1) \, , \\
\label{eq:pairsep_k2_pq}
 k_2 &=& \frac{9}{4} \, .
\end{eqnarray}
It is now clear that the solution $P(\Delta,\tau)$ in our model is solely determined by $k_0=k_0(q^*,\alpha)$ for a given initial condition. The solution $P(\Delta,t)$, with time $t$ rather than the scaled time $\tau$, is then given by an additional rescaling of the time axis. It is therefore apparent that even if it is possible to completely determine $P(\Delta,t)$ by experiment or DNS, this can only lead to the determination of two parameters. It will not be possible to determine all three model parameters $\alpha$, $q^*$ and $B$ from fitting data to this PDF.

\subsubsection{PDF of pair separation for arbitrary $\alpha$}
\label{sec:pairsep_arbitrary_alpha}

While the assumption $\alpha \ll 1$ from the previous section leads to a compact formulation for the time evolution of the PDF of pair separation, it is a very restrictive assumption which basically states that the ``burst'' processes described earlier happen in a more or less continuous fashion, thus suggesting that there may be no bursts.

We shall now present an argument that the PDF evolution of the form \eref{pairsep_PDF_PDE_ks} is valid for all positive $\alpha$, albeit the coefficients $k_{0,1,2}$ will be more complicated in terms of the model parameters $\alpha$, $p$ and $q$.

First note that the sum over all probabilities is constant in time,
\begin{equation}
\label{eq:pairsep_Qn_constintime}
\frac{\partial}{\partial t} \sum_{n=-\infty}^\infty Q_n = 0 \, ,
\end{equation}
as can easily be verified from \eref{pairsep_Qn_evolution}. Substituting the series with infinite number of terms
\begin{eqnarray}
Q_{n\pm 1} &=& \sum_{j=0}^\infty \frac{(\pm 1)^j}{j!} \frac{\partial^j}{\partial n^j} Q_n = \alpha \sum_{j=0}^\infty \frac{(\pm \alpha)^j}{j!} \left(\Delta\frac{\partial}{\partial \Delta} \right)^j \left(\Delta P \right) \, , \\
\label{eq:pairsep_bn-1_expansion}
b_{n-1} &=& B \Delta^{-2/3} \sum_{j=0}^\infty \frac{1}{j!} \left( \frac{2}{3}\alpha \right)^j = B \Delta^{-2/3} e^{2\alpha /3} \, 
\end{eqnarray}
as well as \eref{def_bn} and \eref{pairsep_Qn_P} into \eref{pairsep_Qn_evolution} leads to the following evolution equation for the continuous PDF,
\begin{equation}
\label{eq:pairsep_dtP_inf_series}
\frac{1}{B}\, \frac{\partial P}{\partial t} = (p-q) \Delta^{-2/3} P \left(e^{2\alpha /3}-1\right) + \Delta^{-5/3} \sum_{k=1}^\infty \frac{(-1)^k \, p e^{2\alpha /3} + q}{k!} \, \alpha^k\, \left(\Delta \frac{\partial}{\partial \Delta} \right)^k \, \left( \Delta P \right) \, ,
\end{equation}
which can be cast into the form
\begin{equation}
\label{eq:pairsep_dtP_inf_Cform}
\frac{1}{B} \frac{\partial P}{\partial t} = \frac{\partial}{\partial \Delta} \left[ \sum_{k=0}^\infty C_k\, \Delta^{k+1/3}\, \frac{\partial^k}{\partial \Delta^k} P(\Delta) \right] \, ,
\end{equation}
where the first coefficient is given by
\begin{equation}
 C_0 = 3 \sum_{k=1}^\infty \frac{\alpha^k}{k!} \left[ p \left( -\frac{1}{3} \right)^k + q \left( 1 - \left(\frac{2}{3}\right)^k \right) \right] \, ,
\end{equation}
and the remaining coefficients can be determined from the recursive relation
\begin{equation}
 C_n = \frac{1}{n+1/3} \left\lbrace \sum_{k=n}^\infty \frac{\alpha^k}{k!} \left[ p \sum_{j=0}^{k-n} {k \choose j} (-1)^{k-j} \left(\frac{2}{3}\right)^j {k-j+1 \brace n+1}  + q {k+1 \brace n+1} \right] - C_{n-1} \right\rbrace
\end{equation}
with
\begin{equation}
{n \brace k}
=\sum_{j=1}^k (-1)^{k-j} \frac{j^{n-1}}{(j-1)!(k-j)!}
=\frac{1}{k!}\sum_{j=0}^{k}(-1)^{k-j}{k \choose j} j^n
\end{equation}
being the Stirling numbers of the second kind. The first two coefficients,
\begin{eqnarray}
C_0 &=& - \alpha(p-q) + \frac{\alpha^2}{6}(p+5q) - \frac{\alpha^3}{54}(p-19q) + \frac{\alpha^4}{648}(p+65q) - \cdots \, ,\\
C_1 &=& \frac{\alpha^2}{2}(p+q) - \frac{\alpha^3}{18}(5p-11q) + \frac{\alpha^4}{216}(21p+85q) - \cdots \, ,
\end{eqnarray}
agree with \eref{pairsep_k0_pq} - \eref{pairsep_k2_pq} up to second order in $\alpha$. Note that truncation after any $k$ term in \eref{pairsep_dtP_inf_Cform} preserves the property
\begin{equation}
\label{eq:pairsep_Pint_constintime}
\frac{\partial}{\partial t} \int_0^\infty P(\Delta,t) = 0 \, ,
\end{equation}
which is the continuous-$\Delta$ form of \eref{pairsep_Qn_constintime}.
Furthermore, \eref{pairsep_dtP_inf_Cform} can also be rearranged as follows
\begin{equation}
\label{eq:pairsep_Pevo_infty_series}
\frac{1}{B} \frac{\partial P}{\partial t} = \sum_{m=1}^\infty D_m \, \frac{\partial^m}{\partial \Delta^m} \left[ \Delta^{m-2/3}\,  P(\Delta) \right] \, ,
\end{equation}
where the $D_m$ are given by the recursive relation
\begin{equation}
D_m = C_{m-1} - \sum_{k=m}^\infty D_{k+1} {k \choose m-1} \left(m+\frac{1}{3}\right)^{(k-m+1)}
\end{equation}
with
\begin{equation}
(x)^{(n)} = x(x+1)(x+2)\cdots(x+n-1)=\frac{\Gamma(x+n)}{\Gamma(x)}= \frac{(x+n-1)!}{(x-1)!}
\end{equation}
being the Pochhammer symbol or ``rising factorial''. Again, the first two coefficients,
\begin{eqnarray}
D_1 &=& - \alpha(p-q) - \frac{\alpha^2}{6}(3p-q) - \frac{\alpha^3}{18}(3p-q) - \frac{\alpha^4}{648}(27p-5q) - \cdots \, ,\\
D_2 &=& \frac{\alpha^2}{2}(p+q) + \frac{\alpha^3}{6}(3p-q) + \frac{\alpha^4}{72}(21p+5q) + \cdots \, ,
\end{eqnarray}
agree with \eref{pairsep_k0_pq} - \eref{pairsep_k2_pq} up to second order in $\alpha$ and truncation after any $m$ term in \eref{pairsep_Pevo_infty_series} preserves the property \eref{pairsep_Pint_constintime}.

One can now examine the time evolution of the second moment of $P(\Delta)$. Using \eref{pairsep_Pevo_infty_series} and two integrations by parts, one finds
\begin{eqnarray}
\frac{\d}{\d t} \left\langle \Delta^2 \right\rangle &=& \int_0^\infty \Delta^2\, \frac{\partial P(\Delta)}{\partial t} \, \d \Delta \\
\nonumber
&=& -2 \int_0^\infty D_1\, \Delta^{4/3} P \, \d \Delta + 2 \int_0^\infty D_2\, \Delta^{4/3} P \, \d \Delta \\
\label{eq:pairsep_Pmssq_evo_2ndline}
&~&-2 \int_0^\infty \frac{\partial }{\partial \Delta} \left[ \sum_{m=3}^\infty D_m \, \frac{\partial^{m-3}}{\partial \Delta^{m-3}} \left( \Delta^{m-2/3}  P \right) \right] \, \d \Delta \\
&=& 2 \, (D_2-D_1) \left\langle \Delta^{4/3} \right\rangle \, ,
\end{eqnarray}
provided that all derivatives of $P(\Delta)$ involved in the third integral of \eref{pairsep_Pmssq_evo_2ndline} vanish sufficiently fast for $\Delta \rightarrow \infty$:
\begin{equation}
 \forall u \in \{0,1,\ldots,\infty\}:\, \lim_{\Delta \rightarrow\infty} \Delta^{u+7/3}\, \frac{\partial^u\, P(\Delta)}{\partial \Delta^u} = 0 \, .
\end{equation}

The conclusion to be drawn from this equation is that the complete evolution of $\left\langle \Delta^2 \right\rangle$, and thus $\left\langle \Delta^2 \right\rangle$ itself, depends only on the first two terms in \eref{pairsep_Pevo_infty_series} and therefore, can be determined from a Richardson-type equation such as \eref{pairsep_PDF_PDE_tidy}. Hence, truncating the infinite series of \eref{pairsep_Pevo_infty_series} after the second term leads to an evolution equation for the PDF, which contains the zeroth, first and second derivatives of $P(\Delta)$ and gives the correct first and second moments only. Higher order moments of separation will require higher derivative terms to be included. Of course, the coefficients $k_{0,1,2}$ of \eref{pairsep_PDF_PDE_ks} will be functions of $D_1$ and $D_2$, and thus of the model parameters $B$, $p$, $q$ and $\alpha$. Determining these 4 model parameters from the 3 accessible coefficients $k_{0,1,2}$ is generally not possible.

Note that the coefficients $k_{0,1,2}$ contain infinite series of $m$-terms $\alpha^m/m!$, and thus one can expect that these series will converge for arbitrary $\alpha$. To avoid unnecessary complication of the notation, we shall use the $k_{0,1,2}$ from the previous section, obtained under the assumption $\alpha \ll 1$ for the remainder of this article. Note that this does not change the validity of the results, as they only depend on the finiteness of $k_{0,1,2}$, and not on their dependence on the model parameters.

\subsubsection{Time reversal in the new model}

The caption of \fref{pairsep_Delta} might hint that an exchange of the parameters $p$ and $q$ might be sufficient to account for a possible time asymmetry of pair separation. This, however, is misleading.

The present model does not explicitly contain any dynamic features: all assumptions made can be satisfied by a Gaussian velocity field or a kinematic simulation, which is known not to exhibit the sought time asymmetry \cite{Flohr:2000,Frisch:1999,Sawford:2005}. Hence, there is no reason to expect this model to account for the observed asymmetry.

Instead, if one examines a velocity field which does exhibit the time asymmetry of mean square separation, it would imply that the model describes this situation as two separated cases with {\em two} sets of model parameters, one for forward separation and one for backward separation. It therefore can describe the asymmetry, but not explain it.

Furthermore, one can consider the following mathematical argument. From experimental observations of the PDF of separation, one expects the solution $P(\Delta,t)$ to be ``melting'', i.e. initial sharp peaks should flatten and spread out as time advances (in both the forward and backward case). This behavior is only observed when $k_0 < 0$. However \eref{pairsep_k0_pq} can only be negative if $p^* > q^*$. Hence, assuming that $p^* > q^*$ in the forward case, $k_0>0$ in the backward case (or vice versa).

\subsubsection{Comparison between the PDFs derived by Richardson, GV04 and the present work}
\label{sec:pairsep_model_comparison}

All three models \cite{GV04,Richardson:1926} can be represented by an equation of the form \eref{pairsep_PDF_PDE_ks}, where the time variable has been suitably normalized to satisfy $k_2=9/4$. Then, $k_1$ shows the same dependence on $k_0$ in all three models and the $k_0$ are given by:
\begin{eqnarray}
\label{eq:k0_present}
 \mathrm{present :} \quad k_0 &=& -\frac{3}{2}\,\frac{1-2q^*}{\alpha} + q^* + \frac{1}{4} \\
\label{eq:k0_GV04}
 \mathrm{GV04 :} \quad k_0 &=& - \frac{3}{2}\,\frac{1}{\alpha_\mathrm{GV}} + \frac{1}{4}\\
\label{eq:k0_Rich}
 \mathrm{Richardson :} \quad k_0 &=& - \frac{3}{4} \, (d-1)
\end{eqnarray}
The difference between the Richardson model and the other two is that for a given dimension of the problem (i.e. $d=2$ or $d=3$), the Richardson model predicts $k_0$. Both other models do not predict the value of $k_0$, but instead allow deviations from Richardson's PDF. GV04 have shown that their 2D DNS data is best fitted by $\alpha_\mathrm{GV} \approx 1.3$, whereas Richardson's PDF would predict $\alpha_\mathrm{GV} = 1.5$.

Furthermore, for arbitrary $\alpha$, we have shown in \sref{pairsep_arbitrary_alpha} that higher derivate corrections are necessary for higher separation moments. This could potentially explain deviations from Richardson's law for higher moments, should they be sufficiently resolved in experiments in the future.

\subsubsection{Richardson limit as guideline for expected parameter values for $q$ and $\alpha$}

So far, data has been reasonably fitted with Richardson's form of the pair separation PDF \cite{Berg:2006,GV04,Ott:2000}. One can therefore use the $k_0$ value in the Richardson case as a guideline for the expected parameters $\alpha$ and $q$ (or $q^*$; we shall use the ratio $q/p$ instead of $q^*/p^*$ purely for legibility). By equating \eref{k0_present} and \eref{k0_Rich}, one can find a constraint on the parameters $q$ and $\alpha$ to match the pair separation PDF as introduced by Richardson \cite{GV04, Richardson:1926}:
\begin{equation}
\label{eq:RichLimit_q_of_alpha}
 \frac{q}{p}=\frac{q^*}{p^*} = \frac{1-\alpha \left(\frac{d}{2} - \frac{1}{3} \right)}{1+\alpha \left(\frac{d}{2} + \frac{1}{3} \right)} \, .
\end{equation}
One can invert \eref{RichLimit_q_of_alpha} to find an expression for $\alpha$:
\begin{equation}
\label{eq:RichLimit_alpha_of_q}
 \alpha = 6\, \frac{p-q}{p(3d-2)+q(3d+2)} \, ,
\end{equation}
which leads to the following form for two and three dimensions:
\begin{eqnarray}
 \alpha &=& \frac{3}{2}\, \frac{p-q}{p+2q} \qquad \mathrm{for}\; d=2 \\
 \alpha &=& \frac{6}{7}\, \frac{p-q}{p+\frac{11}{7}q} \qquad \mathrm{for}\; d=3 \, .
\end{eqnarray}
Note that for $q=0$ this matches the values found by GV04 \cite{GV04}.

\subsubsection{Analytical solution for vanishing initial separation $\Delta_0=0$}
\label{sec:pairsep_analy_sol}
 
Equation (34) of GV04 \cite{GV04}, is an exact solution for \eref{pairsep_PDF_PDE_ks}, which we formulated here in terms of $k_0$. It is valid for all three cases discussed here:
\begin{equation}
\label{eq:pairsep_PDF_exact}
P(\Delta,\tau) = \frac{2/3}{\Gamma(3/2-2k_0)} \, \tau^{-3/2} \, \left( \frac{\Delta^{2/3}}{\tau} \right)^{-2 k_0} \, \exp\left[ -\frac{\Delta^{2/3}}{\tau} \right] \, .
\end{equation}
Note that this is only a valid solution when $k_0<0$ as is it satisfied by e.g. Richardson and GV04. $\Gamma(\cdot)$ is the Gamma function.

\begin{figure}[htp]
\centering
\includegraphics{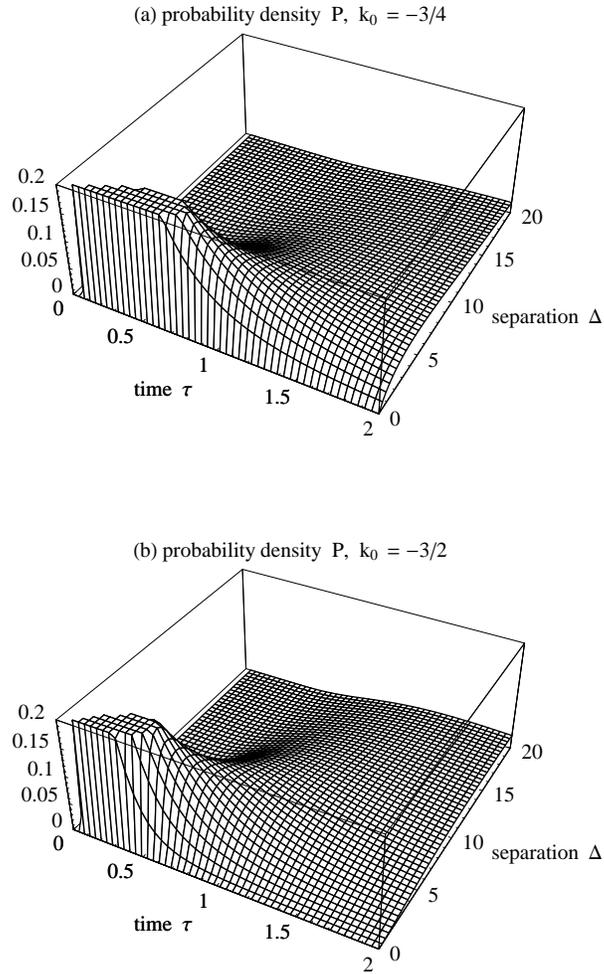}
\caption{PDF of pair separation \eref{pairsep_PDF_exact} with $k_0=-3/4$ (a) and $k_0=-3/2$ (b). Note that the PDF with $k_0=-3/2$ spreads faster with respect to the scaled time $\tau$. See \fref{pairsep_Ptilde} to compare the shape of both PDFs in $\Delta$-direction.}
\label{fig:pairsep_analyticalP}
\end{figure}

This analytical solution (see \fref{pairsep_analyticalP}) has the initial condition $P(\Delta, 0) = 2\delta(\Delta)$ (with $\delta(\cdot)$ being the Dirac delta function) and hence, is not directly applicable to comparison with experiments as the (initial) separation between two physical particles is usually always $>0$. However, if one assumes that after a certain time, the separation process will have ``forgotten'' the value of its initial state $\Delta_0$, for $\tau \gg 0$ \eref{pairsep_PDF_exact} can be compared to experiments with $\Delta_0>0$. This is frequently done in the literature \cite{Berg:2006,GV04,Ott:2000}, and deviations from a fit are usually not discussed with sufficient care.

This solution leads to the familiar $t^3$-law for the mean square separation, irrespective of the parameter $k_0$, which only contributes to the coefficient $G_\tau$:
\begin{equation}
\label{eq:pairsep_msqsep_analytical}
 \left\langle \Delta^2(\tau) \right\rangle = \int_0^\infty \Delta^2\, P(\Delta, \tau) \, \d\Delta = G_\tau\, \tau^3 \, ,
\end{equation}
where
\begin{equation}
 G_\tau = \left(\gamma + \frac{7}{2} \right)\left(\gamma + \frac{5}{2} \right)\left(\gamma + \frac{3}{2} \right) \, ,
\end{equation}
and $\gamma=-2 k_0$ for brevity. More generally, one can find a formula for arbitrary moments of $\Delta$ with $n>-(1+\frac{2}{3}\gamma)$:
\begin{equation}
 \left\langle \Delta^n(\tau) \right\rangle = \frac{\Gamma\left(\gamma + \frac{3}{2} + \frac{3}{2} n \right)}{\Gamma\left(\gamma + \frac{3}{2} \right)}\, \tau^{\frac{3}{2}n} \, .
\end{equation}
Furthermore, normalizing the PDF of separation with its r.m.s. separation leads to a time-independent self-similar PDF \cite{GV04}:
\begin{equation}
\label{eq:pairsep_PDF_PDE_selfsim}
 \tilde{P}(\tilde{\Delta}) = \frac{2 \left( G_\tau^{\gamma + 3/2} \right)^{1/3} }{3\,\Gamma(\gamma + 3/2)} \,  \tilde{\Delta}^{2\gamma/3} \, \exp\left[ - G_\tau^{1/3} \tilde{\Delta}^{2/3} \right] \, ,
\end{equation}
where
\begin{equation}
 \tilde{P}(\tilde{\Delta}) = \sqrt{\left\langle \Delta^2(t) \right\rangle} \, P(\Delta,t) \qquad \mathrm{and} \qquad \tilde{\Delta} = \Delta / \sqrt{\left\langle \Delta^2(t) \right\rangle} \, .
\end{equation}
See \fref{pairsep_Ptilde}.

\begin{figure}[htb]
\centering
\includegraphics{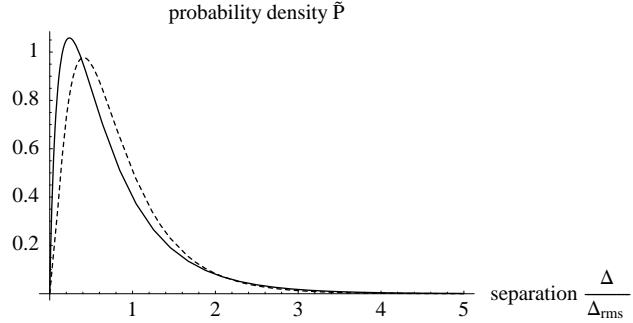}
\caption{
Self-similar PDF $\tilde{P}(\tilde\Delta)$ for vanishing initial separation. $k_0=-3/4$ (solid) and $k_0=-3/2$ (dashed) is what Richardson \cite{Richardson:1926} predicts respectively for 2D and 3D turbulence.}
\label{fig:pairsep_Ptilde}
\end{figure}

The $n$-moments of this self-similar PDF are constants that only depend on $\gamma$:
\begin{equation}
\label{eq:pairsep_rawPtildeMoments}
 \left\langle \tilde{\Delta}^n \right\rangle = \frac{\left\langle \Delta^n \right\rangle}{\left\langle \Delta^2 \right\rangle^\frac{n}{2}} = \frac{\Gamma\left(\gamma + \frac{3}{2} + \frac{3}{2} n \right)}{\Gamma\left(\gamma + \frac{9}{2} \right)^\frac{n}{2}\, \Gamma\left(\gamma + \frac{3}{2} \right)^{1-\frac{n}{2}}} \, .
\end{equation}
Provided one accepts the assumption $\alpha \ll 1$, one can expect these time-independent moments to be a good indicator of the validity of the presented model: the more moments we can fit to the data for a given $\gamma$, the closer this model represents what is going on.
For arbitrary $\alpha$, this relation is only valid for the first and second moment, as has been argued in \sref{pairsep_arbitrary_alpha}.

Since experimental data or data from numerical simulations cannot be obtained for $\Delta_0=0$, we need a means of understanding what happens for finite initial separations.





\section{Two dimensional DNS study}

While the effect of time asymmetry of pair separation has been reported for three dimensional experiments and DNS \cite{Berg:2006,Luethi:2007}, there are no published results about this asymmetry in two dimensional turbulence.

\begin{table}[h!tbp]
\begin{center}
\begin{tabular}{ccccccccc}
\hline\hline\hlinevspace
$N^2$		& $k_f$	& $\beta$	& $\tilde{\omega_0}$ 	& $\d t$ 		& $\nu$ 			& $m_1$ 	& $\alpha$	& $m_2$  \\ \hline \hlinevspace
$3072^2$	& $680$	& $1.00147$	& $0.525$		& $6.1 \times 10^{-5}$	& $1.43 \times 10^{-44}$	& $8$	& $2.5$		& $1$ \\
\hline\hline
\end{tabular} 
\end{center}
\caption{Parameters used in the numerical scheme as described in \cite{GV04}: the number of grid points in the de-aliased scheme $N^2$, forcing wavenumber $k_f$, ratio of forcing wavenumber range $\beta$, magnitude of fixed Fourier component of vorticity $\tilde{\omega_0}$, time increment of temporal integration $\d t$, hyper-viscosity coefficient $\nu$ and its exponent $m_1$, hyper-drag coefficient $\alpha$ and its exponent $m_2$ \cite{GV04}.}
\label{tab:pairsep_dns_parameters}
\end{table}

In this section, we shall present findings from the DNS run with a resolution of $N=3072$. We used the DNS scheme described in \cite{GV04} with the parameters given in \tref{pairsep_dns_parameters} and typical scales given in \tref{pairsep_dns_scales}. Please refer to \cite{GV04} for more details about this DNS, as reiterating the details would lengthen this article unnecessarily. All data were collected using $50000$ particle pairs which were initially placed evenly distributed at random in the entire DNS domain. The distance between pair member particles is fixed at the distance $\Delta_0$ to give a delta function initial state of the pair separation PDF $P(\Delta,0)=\delta(\Delta-\Delta_0)$. The spatial orientation of the separation vector was chosen at random.

\begin{table}[h!tbp]
\begin{center}
\begin{tabular}{l|l}
\hline\hline\hlinevspace
integral length $\L$	& $0.24$ \\
smallest lengthscale in $-5/3$ intertial range $\eta_f = 2\pi/k_f$	& $9.2 \times 10^{-3}$ \\
outer vs. inner length scale $L/\eta_f$	& $26$ \\
r.m.s. velocity fluctuation $u' = \sqrt{\left\langle \vec{u}\cdot\vec{u}\right\rangle /2}$	& $1.1$ \\
integral timescale $T_\L=\L/u'$	& $0.22$ \\
\hline\hline
\end{tabular} 
\end{center}
\caption{All quantities are given in DNS units and will be used to non-dimensionalize where appropriate. Note that $T_\L \approx 3560~\mathrm{timesteps}$ and hence, the DNS covers approximately $11$ integral timescales.}
\label{tab:pairsep_dns_scales}
\end{table}

There are 20 runs in total: 10 for the initial separations,
\begin{equation}
 \Delta_0 \in \left\lbrace \frac{\eta_f}{16}, \frac{\eta_f}{8}, \frac{\eta_f}{4}, \frac{\eta_f}{2}, \eta_f, 2 \eta_f, 4 \eta_f, 8 \eta_f, 16 \eta_f, 32 \eta_f \right\rbrace \, ,
\end{equation}
for both forward and backward separation in time. The particle trajectories are obtained by integrating the DNS velocity field $\vec{u}(\vec{x}(t),t)$,
\begin{equation}
 \vec{x}(t) = \vec{x}_0 + \int_0^t \vec{u}(\vec{x}(t'),t') \, \d t' \, ,
\end{equation} 
using a second order predictor-corrector scheme \cite{BurdenFaires} for $39150$ consecutive time steps\footnote{The number $39150$ arose from hard disk constraints.} of the DNS which are stored on the hard drive. In the backwards case, the sequence of velocity field time frames is inverted and the velocity field negated to account for the reversed particle motion.

\subsection{Validity of model assumptions}
\label{sec:pairsep_validity}

Before comparing the DNS data to the model presented earlier, it is necessary to check which part of the data actually satisfies the assumptions made in the model, namely the presence of a multiscale stagnation point topology which encompasses scales smaller and larger than the separation scale of the particle pairs under observation.

\begin{figure}[h!tbp]
\centering
\includegraphics{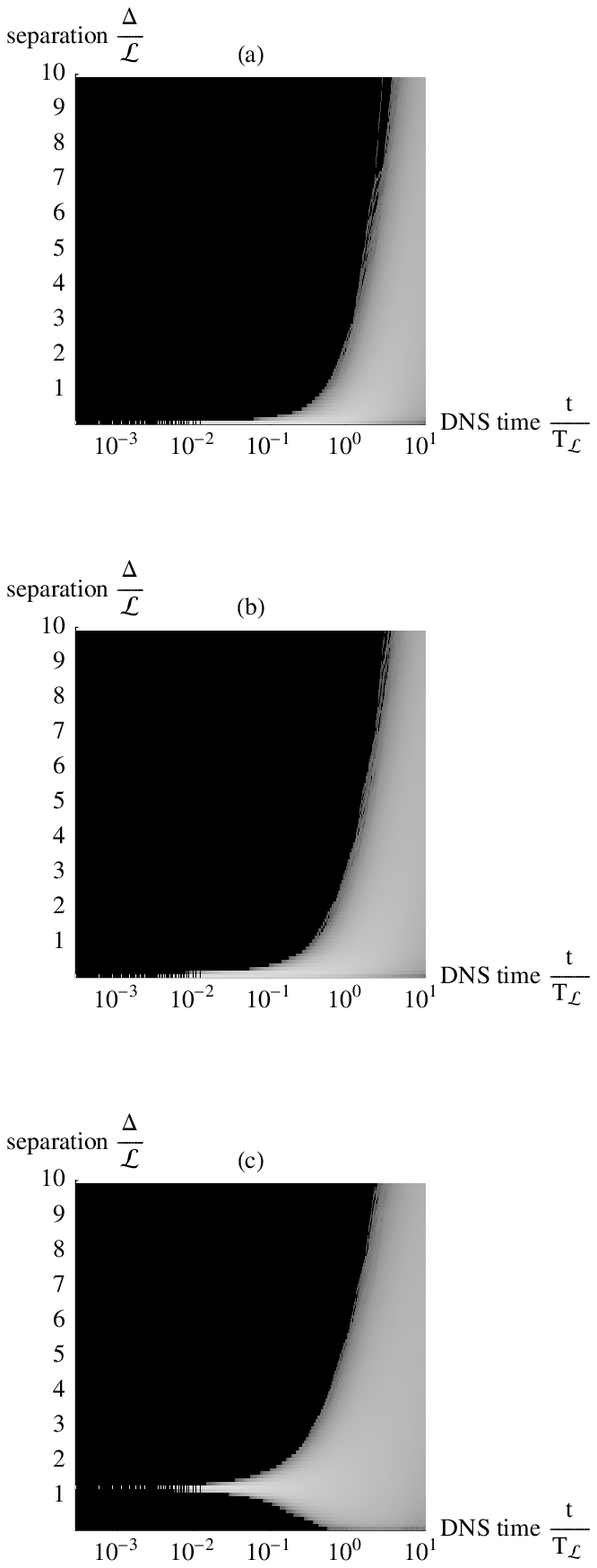}
\caption{Pair separation PDF $P(\Delta,t)$, plotted logarithmically in time. All figures show PDFs from forward separation data, the figures for backwards data look very similar. (a) to (c) have the following initial separations: $\Delta_0 = \frac{\eta_f}{16}, \, \eta_f, \, 32 \eta_f$. Black represents a PDF value of $10^{-6}$ or lower. The probability of a particle pair having a separation of the integral scale $\L$ exceeds values of $10^{-3}$ (gray) at about $t/T_\L \approx 10^{-1}$, see text.}
\label{fig:pairsep_dns_pdf_data_logtime}
\end{figure}

\Fref{pairsep_dns_pdf_data_logtime} shows that the PDF of pair separation is rapidly spreading to larger separations, as would be expected from the exponential tail of the analytical solution \eref{pairsep_PDF_exact}. To quantify the time when many pairs leave the inertial range, we introduce the threshold time $t_{\L}$ which is defined as the time when $P(\L,t)$ crosses an arbitrary threshold for the first time:
\begin{equation}
 P(\L,t_\L) = 10^{-3} \, .
\end{equation}
Note that the order of magnitude of $t_{\L}$ and its dependency on $\Delta_0$ does not change significantly if a different threshold is chosen between $10^{-5}$ and $10^{-2}$. The threshold times for all given $\Delta_0$ show the expected decreasing trend in \fref{pairsep_dns_threshold_time}. For comparison, the Batchelor time,
\begin{equation}
 t_B = \left( \frac{\Delta_0^2}{\epsilon} \right)^{1/3} \, ,
\end{equation}
indicating the end of the ballistic regime \cite{Berg:2006}, is shown in the same figure. The dissipation rate $\epsilon$ has been estimated from the constant energy flux of the DNS velocity field, as described in \cite{GV04}.

\begin{figure}[h!tbp]
\centering
\includegraphics{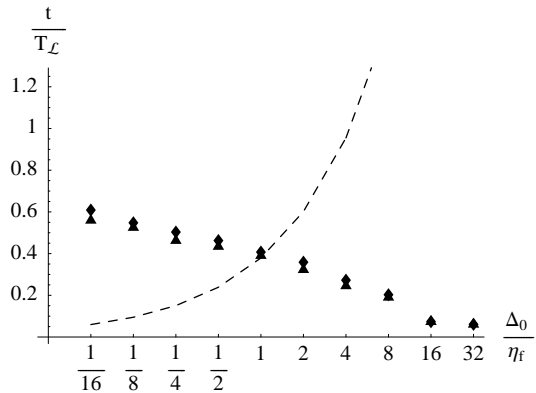}
\caption{Threshold time $t_{\L}$ for all initial separations $\Delta_0$. Forward separation data is shown as triangles ($\blacktriangle$), backwards data as diamonds ($\blacklozenge$). For comparison, the Batchelor time $t_B$ is represented by the dashed line.}
\label{fig:pairsep_dns_threshold_time}
\end{figure}

Summarizing, $t_{\L}$ marks the time at which approximately $0.1\%$- $1\%$ of pairs have a separation larger than the integral scale $\L$ and therefore have left the inertial range. The Batchelor time $t_B$ marks the time when initial separation effects are assumed to have ceased \cite{Batchelor:1950,Ott:2000} and after which the Richardson $\Delta \propto t^3$ law is expected. Also, separations below the forcing length $\eta_f$ are outside the inertial range.

Therefore, the presented model is strictly only valid for data with
\begin{itemize}
 \item $\eta_f < \Delta_0 < \L$, i.e. initial separation within the inertial range, and
 \item $t < t_{\L}$, i.e. enough pairs remain within the inertial range.
\end{itemize}
Furthermore, the Richardson $t^3$ law can only be expected for $t>t_B$. In conclusion, the resolution of $N=3072$ does not provide an inertial range that is wide enough to be able to observe the Richardson $t^3$ law \cite{Berg:2006,Boffetta:2002,Yeung:2002}. When comparing the data with the present model, it is therefore necessary to discuss the influence of these adverse effects in sufficient detail.

\label{sec:pairsep_suitability_on_analy_sol}
The same applies when one wishes to compare the analytical solution \eref{pairsep_PDF_exact} with no initial separation to DNS data: At the times the fit is performed, many pairs have left the inertial range already ($t_\mathrm{cutoff} \gg t_\L$) and hence, a basic model assumption is violated. To reflect this effect, the model would need to be extended by e.g. modelling the particle behavior similar to Brownian motion for pairs with $\Delta > \L$. However, this extension goes beyond the scope of this work, whose primary objective is to investigate the multiscale stagnation point topology within the inertial range.

\subsection{Mean square separation and parameter fit}
\label{sec:pairsep_msqs_fit}

Despite the conclusion in \sref{pairsep_validity} that it is not expected to observe a $t^3$ law in the given DNS data, \fref{pairsep_dns_msqsep_data} shows a slope that is approximately close to $t^3$ for times $t \in [t_B,t_\L]$ for the runs with initial separation smaller than the forcing length: $\Delta_0 \leq \eta_f$.

\begin{figure}[h!tbp]
\centering
\includegraphics{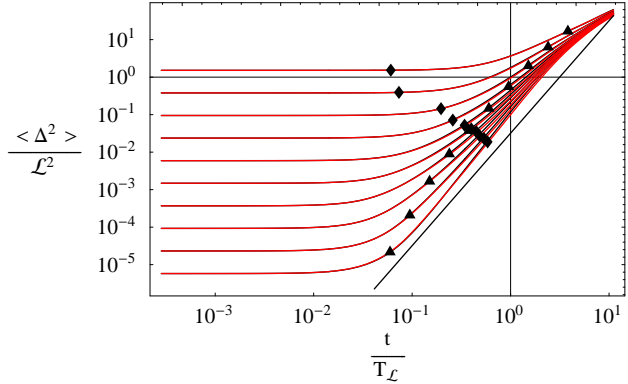}
\caption{Mean square separation from the DNS for all 10 initial separations. Forward separation is shown in black, backward in red. The Batchelor time $t_B$ is indicated by triangles ($\blacktriangle$) and the threshold time $t_\L$ by diamonds ($\blacklozenge$). The straight line is $\propto t^3$. The slope of the data is approximately $\propto t^3$ in the intervals $[t_B,t_\L]$, although there is a noteable deviation for times just larger than $t_B$. The $5^\mathrm{th}$ graph from the bottom has initial separation $\Delta_0=\eta_f$.}
\label{fig:pairsep_dns_msqsep_data}
\end{figure}

According to \sref{pairsep_validity}, this behavior was only to be expected for $\Delta_0 > \eta_f$. However, as mentioned previously, the DNS resolution is currently not large enough to accommodate $t_B < t_\L$ for initial separations larger than $\eta_f$.

Nonetheless, the approximate slope $\propto t^3$ is quite meaningless, as for times $t< t_\L$, the mean square separation is still influenced by the finite initial separation. Thus, let us compare the model's prediction for $\left\langle \Delta^2\right\rangle$ with finite initial separation $\Delta_0>0$ with the data and try to fit the model parameters.

The model has been integrated for all 10 initial separations and for 3 different values of $k_0$, namely $k_0 \in \left\lbrace -0.65, -0.75, -0.85 \right\rbrace $. The parameter $\tau/t$ can be easily adjusted after integration of the model. This parameter has been manually adjusted to give a qualitative fit to the DNS data for the three values of $k_0$ mentioned previously. The two best fitting DNS data sets are shown in \fref{pairsep_dns_mssq_bestfit} with $\Delta_0 \in \left\lbrace \eta_f/4, \eta_f/2\right\rbrace $ for the best fit value of the time scaling parameter, $\tau/t \approx 1.2$.

For initial separations in our data set which are smaller or larger than the best fit ones pointed out, the fit with these parameter values become less acceptable. Using different parameters improves the situation, but does not lead to a fit of the same quality as shown in \fref{pairsep_dns_mssq_bestfit}.

Hence, we conclude that one set of parameters is not sufficient to fit all data for varying initial separations.

\begin{figure}[h!tbp]
\centering
\includegraphics{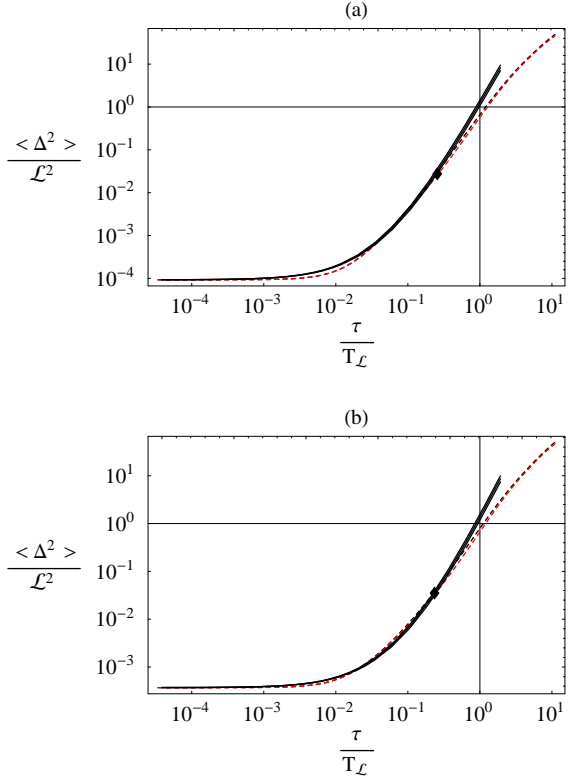}
\caption{ Best fit of numerically integrated mean square separation $\left\langle \Delta^2\right\rangle$ to DNS data with initial separation $\Delta_0=\eta_f/4$ (a) and $\Delta_0=\eta_f/2$ (b). The DNS data is shown in dashed lines, black (forward) and red (backward). The model data is given in black for three values of $k_0$: $-0.85, -0.75, -0.65$ (top to bottom). The lines are very close together, indicating a weak dependence of $\left\langle \Delta^2\right\rangle$ on $k_0$. The diamonds ($\blacklozenge$) demark the threshold time $t_{\L}$, which is the largest time for which one can expect agreement of the model with the data.}
\label{fig:pairsep_dns_mssq_bestfit}
\end{figure}

From \fref{pairsep_dns_mssq_bestfit}, it is immediately apparent that the variation of $k_0$ only has a small impact on the variation of the curve and therefore, it is not possible to estimate $k_0$ to a very high accuracy from this kind of fit. Hence, while Richardson's value of $k_0$ is compatible with the present data, it can unfortunately not be uniquely confirmed. Nevertheless, by applying Ockham's razor, we shall assume Richardson's value of $k_0=-0.75$ to be a reasonable fit for the remainder of this work. 


\subsection{Dependence on initial separation}
\label{sec:pairsep_dependence_initsep}

The previous section highlights that the present model can not be fitted to the DNS data for all given initial separations. The validity of the model is thus clearly dependent on the initial separation, which is a further manifestation of the limited range of scales that the model operates on: In the DNS, too many particle pairs leave the range of multi-scale stagnation point distances too quickly and therefore, $\left\langle \Delta^2\right\rangle$ falls below the value expected from the present model. This effect is stronger for larger initial separations (see \fref{pairsep_dns_mssq_simpleform}). Hence, additional care needs to be applied when fitting data to the model, as the DNS data will deviate from the model for large times, as seen in \fref{pairsep_dns_mssq_simpleform}.

\begin{figure}[h!tbp]
\centering
\includegraphics{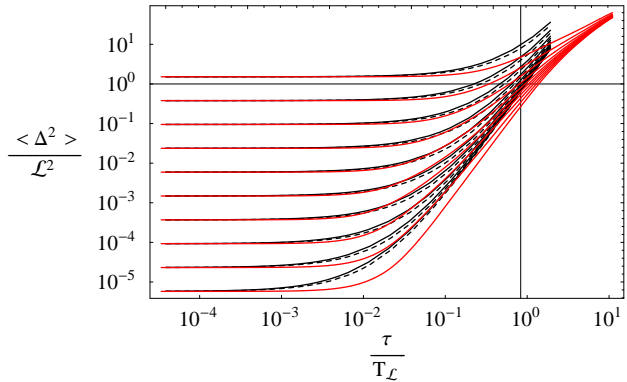}
\caption{ Numerically integrated mean square separation $\left\langle \Delta^2\right\rangle$ (solid black) compared to the simple approximation \eref{pairsep_mssq_simple_approx} (dashed black) and DNS data (solid red) for all 10 initial separations, using the best fit values $\tau/t=1.2$ and $k_0=-0.75$.}
\label{fig:pairsep_dns_mssq_simpleform}
\end{figure}

For comparison, a simple functional form for the mean square separation with finite initial separation has been previously suggested to be used as a fitting law \cite{GV04,Ott:2000}:
\begin{equation}
\label{eq:pairsep_mssq_simple_approx}
 \left\langle \Delta^2 \right\rangle = \left( [G_\Delta\, \epsilon]^{1/3}\, t + \Delta_0^{2/3} \right)^3 = \left( G_\tau^{1/3}\, \tau + \Delta_0^{2/3} \right)^3 \, .
\end{equation}
This is shown in \fref{pairsep_dns_mssq_simpleform}, along with the present model and the DNS data. It easily satisfies the two given limits $\left\langle \Delta^2 \right\rangle =\Delta_0$ for $t=0$ and Richardson's law \eref{pairsep_msqsep_analytical} for $t \rightarrow \infty$. In general, the present model and this simple approximation are more similar to each other than each compared with the DNS data. For the best fit cases with initial separation $\Delta_0 \in \left\lbrace \eta_f/4, \eta_f/2\right\rbrace $ ($3^\mathrm{rd}$ and $4^\mathrm{th}$ from the bottom in \fref{pairsep_dns_mssq_simpleform}), the present model is a better fit than the given approximation \eref{pairsep_mssq_simple_approx}.

Thus, this approximation correctly traces the qualitative features of the present model and hence, provides a useful analytical approximation which can be used as an estimate. However, its use for accurate data fitting remains questionable.

\subsection{Time asymmetry in two dimensional turbulence}
\label{sec:pairsep_time_asymm_2d}

\begin{figure}[h!tbp]
\centering
\includegraphics{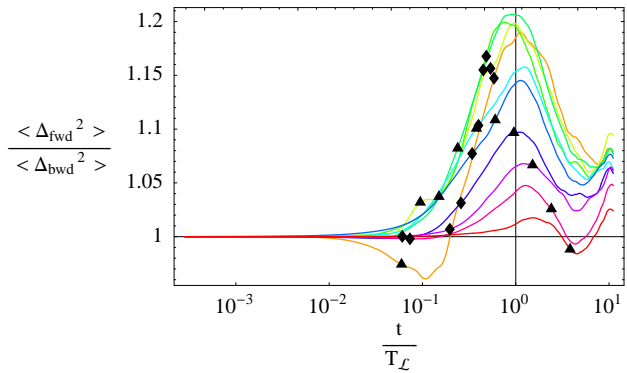}
\caption{ Ratio of mean square separations $\left\langle \Delta_\mathrm{fwd}^2 \right\rangle/\left\langle \Delta_\mathrm{bwd}^2 \right\rangle$ for all initial separations from the DNS data. The lines show the largest initial separation to the smallest through the color spectrum red, blue, green, yellow, orange. If color is not available, this corresponds roughly to bottom to top order at $t \approx T_\L$. The Batchelor time $t_B$ is indicated by triangles ($\blacktriangle$) and the threshold time $t_\L$ by diamonds ($\blacklozenge$). Note that the ``best fit'' cases with initial separation $\Delta_0 \in \left\lbrace \eta_f/4, \eta_f/2\right\rbrace $ are two (green) curves near the highest achieved ratio.}
\label{fig:pairsep_dns_mssq_ratio}
\end{figure}

Looking at \fref{pairsep_dns_msqsep_data}, it seems that there is no significant difference in mean square separation between the forward and backward case in the investigated two dimensional stationary turbulence DNS.

However, the logarithmic scale is deceiving, as the plot of the ratio $\left\langle \Delta_\mathrm{fwd}^2 \right\rangle/\left\langle \Delta_\mathrm{bwd}^2 \right\rangle$ in \fref{pairsep_dns_mssq_ratio} shows. It indicates that the the mean square separation in the forward case is up to $22\%$ larger than the backward case.

\begin{figure}[h!tbp]
\centering
\includegraphics{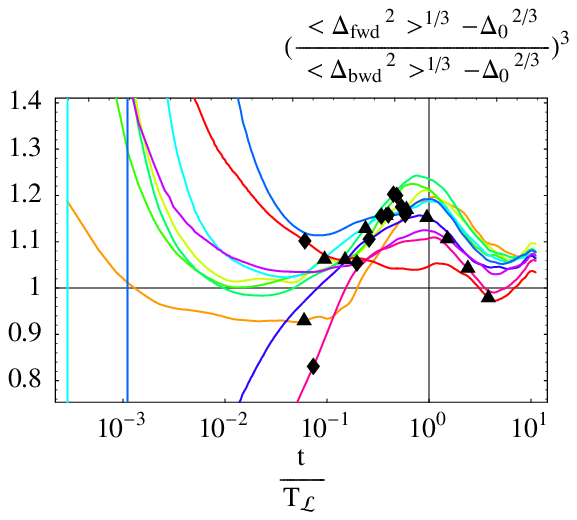}
\caption{Ratio of Richardson's constants $g_f/g_b$ from the approximation \eref{pairsep_mssq_simple_approx} for all initial separations from the DNS data. The lines show the largest initial separation to the smallest through the color spectrum red, blue, green, yellow, orange. If color is not available, this corresponds roughly to bottom to top order at $t \approx T_\L$. The Batchelor time $t_B$ is indicated by triangles ($\blacktriangle$) and the threshold time $t_\L$ by diamonds ($\blacklozenge$).}
\label{fig:pairsep_dns_mssq_ratio2}
\end{figure}

The time asymmetry for a 3D PTV\footnote{Particle Tracking Velocimetry} experiment with $\mathrm{Re}_\lambda \approx 170$ has been documented in the literature by giving the ratio of the Richardson constants $g_f=G_{\Delta,\mathrm{fwd}}$ and $g_b=G_{\Delta,\mathrm{bwd}}$ from a fit of \eref{pairsep_mssq_simple_approx} \cite{Berg:2006}. In this 3D case, the observed ratio was $g_b/g_f \approx 2.1$, i.e. the backward case separated stronger than the forward case. This ratio has been confirmed in the same work by a DNS with $\mathrm{Re}_\lambda \approx 280$. 

To be able to draw a more direct comparison to the 2D case, we shall plot the same ratio in \fref{pairsep_dns_mssq_ratio2}, obtained from \eref{pairsep_mssq_simple_approx}, over time:
\begin{equation}
 \frac{g_f}{g_b} = \left( \frac{\left\langle \Delta_\mathrm{fwd}^2 \right\rangle^{1/3} - \Delta_0^{2/3}}{\left\langle \Delta_\mathrm{bwd}^2 \right\rangle^{1/3} - \Delta_0^{2/3}} \right)^3 \, .
\end{equation}
This relation is useful, since it takes the qualitative dependence of Richardson's law on the initial separation $\Delta_0$ into account, as described in \sref{pairsep_dependence_initsep}.
\Fref{pairsep_dns_mssq_ratio2} shows that this ratio is larger than unity for most initial separations and times. Furthermore, with the exception of the $\Delta_0=\eta_f/16$ dataset, around (and between) the critical times $t_B$ and $t_\L$, the ratio is between $1.05$ and $1.25$. Hence, the best estimate from \fref{pairsep_dns_mssq_ratio} and \fref{pairsep_dns_mssq_ratio2} for the ratio of Richardson's constants in this 2D DNS is:
\begin{equation}
 \frac{g_f}{g_b} = (1.15 \pm 0.10) \, .
\end{equation}
Note that in 2D the forward separation is stronger while the effect is opposite in the 3D experiment, where backward separation dominates. However, the effect in 2D with $\sim 15\%$ difference is not as prominant as in 3D ($\sim 100\%$). The figures \fref{pairsep_dns_mssq_ratio} and \fref{pairsep_dns_mssq_ratio2} suggest that the ratio is different from unity. However, given the uncertainty of the ratio, one cannot conclude this for certain from this kind of fit.

\subsection{Richardson's constant and improvement of time asymmetry fit}
\label{sec:pairsep_richardson_const}

While it was possible to estimate the ratio of Richardson's constants from the time series of mean square separation $\left\langle \Delta^2(t) \right\rangle$ for both the forward and backward case, it is well known that obtaining Richardson's constant itself from this kind of data is much more problematic \cite{Boffetta:2002,Nicolleau:2004}. Alternative suggestions of extracting Richardson's constant include statstics of doubling times \cite{Boffetta:2002} and investigation of mean diffusivity depending on mean separation \cite{Nicolleau:2004}.

We will follow the latter approach. Firstly, note that Richardon's constant $G_\Delta$ is usually defined in terms of the $t^3$ law:
\begin{equation}
\label{eq:pairsep_Richardsons_law}
 \left\langle \Delta^2(t) \right\rangle = G_\Delta\, \epsilon\, t^3 \, ,
\end{equation}
where $\epsilon$ is the mean energy dissipation of the flow. From this, one can easily derive the relation
\begin{equation}
\label{eq:pairsep_diffusivity_fit}
 \frac{\d}{\d t} \left\langle \Delta^2(t) \right\rangle = 3\, \left( G_\Delta\, \epsilon \right)^{1/3}\, \left\langle \Delta^2(t) \right\rangle^{2/3} \, ,
\end{equation}
which is valid for all times if one assumes that the Richardson law holds.
This relation also agrees with \eref{pairsep_mssq_simple_approx}, taking the qualitative dependence on initial separation into account.

\begin{figure}[h!tbp]
\centering
\includegraphics{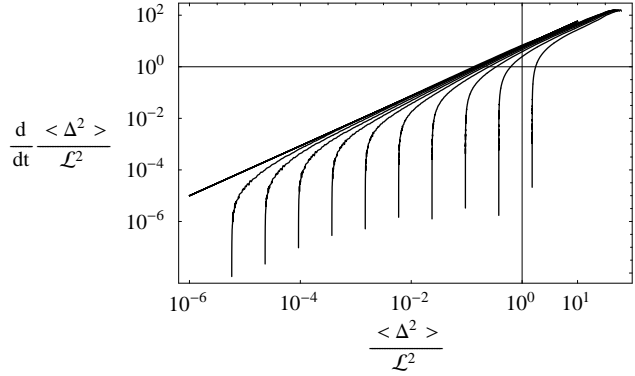}
\caption{Diffusivity $\d/\d t \left\langle \Delta^2(t) \right\rangle$ vs. mean square separation $\left\langle \Delta^2(t) \right\rangle$ for all initial separations from the forward DNS data (smallest to largest left to right). The expected slope of $2/3$ as predicted by \eref{pairsep_diffusivity_fit} cannot be observed. Instead the best fit (straight line) suggests the exponent $1$ indicating exponential growth.}
\label{fig:pairsep_dns_dtmsqsep_vs_msqsep}
\end{figure}

To analyze the DNS data for $\left\langle \Delta^2(t) \right\rangle$, the value of the time derivative is obtained by using a finite difference scheme of second order. This data is given in logarithmically equidistant bins and therefore, the scheme needs to account for variable distances between the data points:
\begin{equation}
 \frac{\d f}{\d t} = \frac{1}{\d t_{-1} + \d t_1} \, \left[ \frac{\d t_{-1}}{\d t_1}\, f(t+\d t_1) - \left( \frac{\d t_{-1}}{\d t_1} - \frac{\d t_1}{\d t_{-1}} \right)\, f(t) - \frac{\d t_1}{\d t_{-1}}\, f(t-\d t_{-1}) \right] + \O{\d t^3} \, ,
\end{equation}
where $f(t)$ is an arbitrary function, $\d t_{-1}$ is the distance to the previous timestep and $\d t_1$ is the distance to the next timestep.

However, it was not possible to observe relation \eref{pairsep_diffusivity_fit} in the DNS data as can be seen in \fref{pairsep_dns_dtmsqsep_vs_msqsep}. Instead evidence for exponential growth is found. Nicolleau and Yu \cite{Nicolleau:2004} found similar behavior for small separations in a 3D kinematic simulation with a large inertial range. While they also observed a region $\propto \left\langle \Delta^2 \right\rangle^{2/3}$, we suspect that the absence of this observation in \fref{pairsep_dns_dtmsqsep_vs_msqsep} is due to the limited width of the inertial range.

Note that independently of the slope of \fref{pairsep_dns_dtmsqsep_vs_msqsep}, the ratio of change of mean square separation also gives the ratio of Richardson's constants:
\begin{equation}
\label{eq:pairsep_timeder_msqsep_ratio}
 \frac{\d/\d t\, \left\langle \Delta_\mathrm{fwd}^2(t) \right\rangle}{\d/\d t\, \left\langle \Delta_\mathrm{bwd}^2(t) \right\rangle} = \frac{ G_{\Delta,\mathrm{fwd}}^{1/3}\, \left\langle \Delta_\mathrm{fwd}^2(t) \right\rangle^{2/3} }{ G_{\Delta,\mathrm{bwd}}^{1/3}\, \left\langle \Delta_\mathrm{bwd}^2(t) \right\rangle^{2/3} } = \frac{ G_{\Delta,\mathrm{fwd}}}{ G_{\Delta,\mathrm{bwd}}} = \frac{g_f}{g_b}\, ,
\end{equation}
where we used \eref{pairsep_Richardsons_law} in the second last equality. Taking finite initial separations into account by using  \eref{pairsep_mssq_simple_approx} gives the same result for large enough times.

\begin{figure}[h!tbp]
\centering
\includegraphics{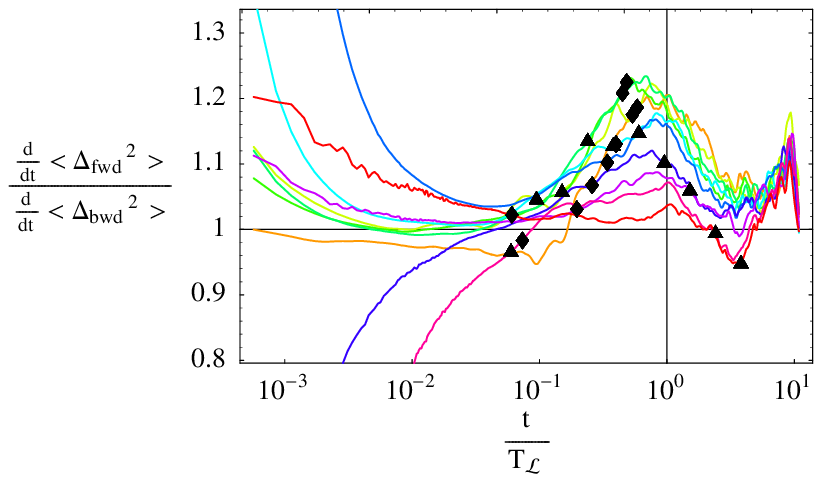}
\caption{ Ratio of Richardson's constants $g_f/g_b$ as given by \eref{pairsep_timeder_msqsep_ratio} for all initial separations from the DNS data. The lines show the largest initial separation to the smallest through the color spectrum red, blue, green, yellow, orange. If color is not available, this corresponds roughly to bottom to top order at $t \approx T_\L$. The Batchelor time $t_B$ is indicated by triangles ($\blacktriangle$) and the threshold time $t_\L$ by diamonds ($\blacklozenge$).}
\label{fig:pairsep_dns_dtmssq_ratio}
\end{figure}

Plotting this ratio using the DNS data in \fref{pairsep_dns_dtmssq_ratio} looks remarkably like \fref{pairsep_dns_mssq_ratio2} and is also consistent with the value of
\begin{equation}
\label{eq:pairsep_richardson_const_ratio}
 \frac{g_f}{g_b} = (1.15 \pm 0.10) \, .
\end{equation}
found earlier. Despite the fact that \fref{pairsep_dns_dtmsqsep_vs_msqsep} suggests non-Richardson exponential growth of $\left\langle \Delta^2(t) \right\rangle$ due to the limited power-law range of length scales, this still is evidence that the forward separation is happening faster than the backward separation.

\subsubsection{Conditioning on separation scale}
\label{sec:pairsep_richardson_const_conditioned}

Nicolleau and Yu \cite{Nicolleau:2004} have demonstrated that it is possible to have exponential and Richardson like growth in the same flow. Our suspicion is that the Richardson like behavior is still present in the DNS flow, but overshadowed by contaminations from scales outside the inertial range, which are present in every average over all Lagrangian trajectories.

Going back to Richardson's original concept that the diffusivity is purely scale dependent \cite{Richardson:1926}, we are going to subject \eref{pairsep_diffusivity_fit} to conditioning on separation scales $\Delta$:
\begin{equation}
\label{eq:pairsep_diffusivity_fit_conditioned}
 \left\langle \frac{\d \Delta^2}{\d t} | \Delta \right\rangle = 3\, \left( G_\Delta\, \epsilon \right)^{1/3}\, \left\langle \Delta^2 | \Delta \right\rangle^{2/3} = 3\, \left( G_\Delta\, \epsilon \right)^{1/3}\, \Delta^{4/3} \, .
\end{equation}
Note that this average is entirely different from the previous ones. 
While in previous sections of this article, the averaging was carried out by sampling the set of Lagrangian trajectories for each timestep in the DNS (i.e. ensemble average, conditioned on DNS time $t$), it is now completely independent of time and instead averages over all particles pairs that have a certain separation $\Delta$ at any given time. Hence, temporal information is lost for averages of this kind, but the conditioning on scales ensures that scale contamination is completely eliminated.

Since this approach does not directly use time series data, the finite difference approach of obtaining the time derivative is not practical. Instead, let us consider the relative motion of particle $1$ and $2$ in a pair:
\begin{equation}
\label{eq:pairsep_dt_Delta}
 \frac{\d}{\d t}\, \Delta = \frac{\d}{\d t}\, \sqrt{(\vec{x_1}-\vec{x_2})^2} = \frac{(\vec{x_1}-\vec{x_2}) \cdot (\vec{u_1}-\vec{u_2})}{\Delta} = (\vec{u_1}-\vec{u_2}) \cdot \vec{\hat\Delta} \equiv u_\Delta \, ,
\end{equation}
where $\vec{x_i}$ and $\vec{u_i}$ are particle position and velocity and $\vec{\hat\Delta} = (\vec{x_1}-\vec{x_2})/\Delta$ is the unit vector of particle separation. $u_\Delta$ is then the longitudinal velocity increment\footnote{Note that this longitudinal velocity increment in not the same as the one generally used in e.g. Kolmogoroff's $4/5$ law. The latter velocity increment is an average over all positions or ensembles at a particular time, whereas the velocity increment defined in \eref{pairsep_dt_Delta} is an average over specific particle pairs at different times. A comparison of dimensions could lead to a suspicion that the two could have the same value. However, this is unlikely since Kolmogoroff's $4/5$ law has opposite signs in 2D and 3D, while $u_\Delta$ is positive in both 2D and 3D.} of a particle pair, which can be calculated from the velocity field at every instance. Thus, we obtain the change of mean square separation from
\begin{equation}
 \left\langle \frac{\d \Delta^2}{\d t} \right\rangle = 2\, \left\langle \Delta\, u_\Delta \right\rangle \, .
\end{equation}
We can now without problem introduce conditioning on separation:
\begin{equation}
\label{eq:pairsep_dt_Delta_conditioned}
 \left\langle \frac{\d \Delta^2}{\d t} | \Delta \right\rangle = 2\, \left\langle \Delta\, u_\Delta | \Delta \right\rangle = 2\,\Delta\, \left\langle u_\Delta | \Delta \right\rangle \, .
\end{equation}
Note that the last equality holds for conditioning using infinitesimal $\Delta$ bins, but for the purposes of obtaining data using finite width separation bins, more accuracy is obtained by using the second expression. Thus, combining \eref{pairsep_diffusivity_fit_conditioned} and \eref{pairsep_dt_Delta_conditioned} we arrive at a relation with can be tested against the DNS data:
\begin{equation}
\label{eq:pairsep_diffusivity_fit_cond_final}
 \left\langle \Delta\, u_\Delta | \Delta \right\rangle = \frac{3}{2}\, \left( G_\Delta\, \epsilon \right)^{1/3}\, \Delta^{4/3} \, .
\end{equation}

\begin{figure}[h!tbp]
\centering
\includegraphics{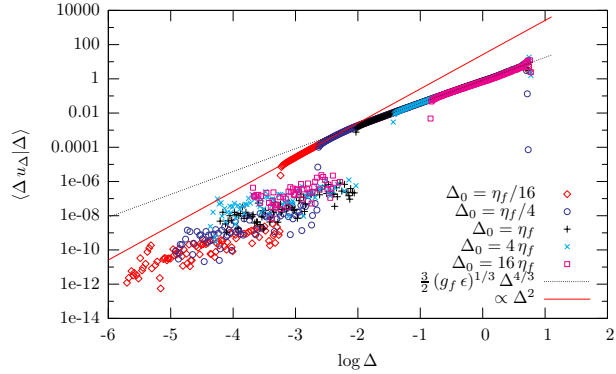}
\caption{$\left\langle \Delta\, u_\Delta | \Delta \right\rangle$ vs. pair separation $\Delta$ for some initial separations from the forward DNS data. The expected slope of $4/3$ as predicted by \eref{pairsep_diffusivity_fit_cond_final} can clearly be observed for at least the inertial range ($-2 \approx \log\eta_f < \log\Delta < \log\L \approx -0.6$) with $g_f = (1.066 \pm 0.020)$. For smaller separations, the expected exponential growth for ballistic separation with slope $\propto \Delta^2$ is also observed.
Note that separations $\Delta < \Delta_0$ have only a small number of pairs contributing to the statistics and therefore, exhibit a large scatter.}
\label{fig:pairsep_dns_dtmssq_delta_fwd}
\end{figure}

\begin{figure}[h!tbp]
\centering
\includegraphics{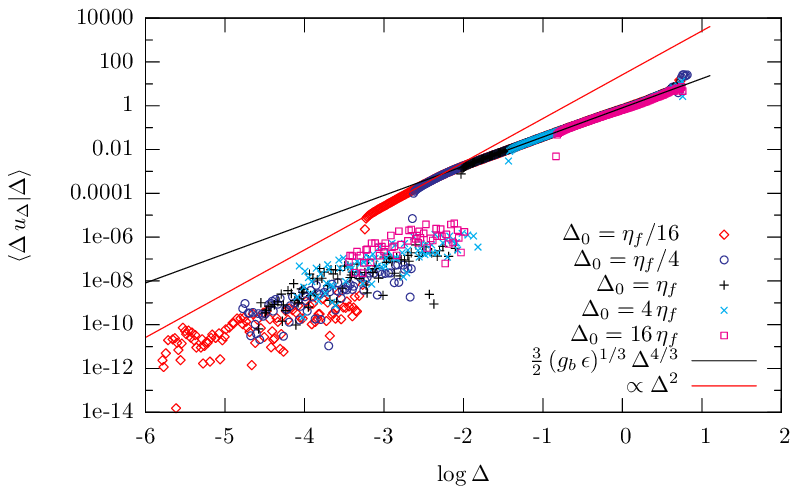}
\caption{$\left\langle \Delta\, u_\Delta | \Delta \right\rangle$ vs. pair separation $\Delta$ for some initial separations from the backward DNS data. The expected slope of $4/3$ as predicted by \eref{pairsep_diffusivity_fit_cond_final} can clearly be observed for at least the inertial range ($-2 \approx \log\eta_f < \log\Delta < \log\L \approx -0.6$) with $g_b = (0.999 \pm 0.007)$. For smaller separations, the expected exponential growth for ballistic separation with slope $\propto \Delta^2$ is also observed.
Note that separations $\Delta < \Delta_0$ have only a small number of pairs contributing to the statistics and therefore, exhibit a large scatter.}
\label{fig:pairsep_dns_dtmssq_delta_bwd}
\end{figure}

\begin{table}[h!tbp]
\begin{center}
\begin{tabular}{cccc}
\hline\hline\hlinevspace
$\Delta_0$	& $g_f$			& $g_b$ 		& $g_f/g_b$  \\ \hline \hlinevspace
$\eta_f/16$	& $1.061\pm 0.004$	& $1.005\pm 0.002$	& $1.06 \pm 0.01$ \\
$\eta_f/8$	& $1.084\pm 0.004$	& $0.996\pm 0.003$	& $1.09 \pm 0.01$ \\
$\eta_f/4$	& $1.080\pm 0.004$	& $0.999\pm 0.002$	& $1.08 \pm 0.01$ \\
$\eta_f/2$	& $1.053\pm 0.004$	& $0.996\pm 0.002$	& $1.06 \pm 0.01$ \\
$\eta_f$	& $1.050\pm 0.004$	& $1.001\pm 0.003$	& $1.05 \pm 0.01$ \\ \hline \hlinevspace
combined	& $1.066\pm 0.020$	& $0.999\pm 0.007$	& $1.07 \pm 0.03$ \\
\hline\hline
\end{tabular} 
\end{center}
\caption{Results from fit of \eref{pairsep_diffusivity_fit_cond_final} to the DNS data. Given are the best fit values of Richardson's constants for forward and backward DNS data and confidence intervals as found by \texttt{gnuplot 4.0} for separations within the inertial range ($-2 \approx \log\eta_f < \log\Delta < \log\L \approx -0.6$). Data with $\Delta_0 > \eta_f$ did not have good enough statistics within the inertial range to be considered for this fit. The uncertainty given for the individual fits is the asymptotic standard error given by gnuplot and the uncertainty given for the combined (averaged) values is the standard deviation $\sigma_{n-1}$ of the distribution of values above plus the average error above.}
\label{tab:pairsep_dns_fit_richardson_const}
\end{table}

The forward and backward DNS data is shown in \fref{pairsep_dns_dtmssq_delta_fwd} and \fref{pairsep_dns_dtmssq_delta_bwd}. Both demonstrate the validity of \eref{pairsep_diffusivity_fit_cond_final} within the inertial range. It seems that this behavior is extended for separations even larger than $\L$, while for $\Delta<\eta_f$ the exponential separation typical of ballistic separation is observed.

As expected, the relation \eref{pairsep_diffusivity_fit_cond_final} is independent of initial separation which manifests itself in a good collapse of all datasets, proving it a useful tool to circumvent the problem of dependence on finite initial separations.

The best fits of Richardson's constants from \eref{pairsep_diffusivity_fit_cond_final} for both forward and backward separation are given in \tref{pairsep_dns_fit_richardson_const}. The overall estimate for Richardson's constant in this 2D turbulence is $g_f = (1.066\pm 0.020)$ for forward separation and $g_b = (0.999\pm 0.007)$ for backward separation. This leads to a ratio of \begin{equation}
\label{eq:pairsep_richardson_const_ratio_2}
 \frac{g_f}{g_b} = (1.07 \pm 0.03)
\end{equation}
which is compatible with the earlier value \eref{pairsep_richardson_const_ratio}.

A further consistency check can be performed by fitting a constant to the ratio of Richardson constants as obtained from \eref{pairsep_diffusivity_fit_cond_final}:
\begin{equation}
\label{eq:pairsep_diffusivity_ratio_fit_cond_final}
 \frac{\left\langle \Delta\, u_\Delta | \Delta \right\rangle_\mathrm{fwd}}{\left\langle \Delta\, u_\Delta | \Delta \right\rangle_\mathrm{bwd}} = \left( \frac{G_{\Delta,\mathrm{fwd}}}{G_{\Delta,\mathrm{bwd}}} \right)^{1/3}  = \left( \frac{g_f}{g_b} \right)^{1/3} 
 \qquad \Rightarrow \qquad \frac{g_f}{g_b} = \left( \frac{\left\langle \Delta\, u_\Delta | \Delta \right\rangle_\mathrm{fwd}}{\left\langle \Delta\, u_\Delta | \Delta \right\rangle_\mathrm{bwd}} \right)^3 \, .
\end{equation}

\begin{figure}[h!tbp]
\centering
\includegraphics{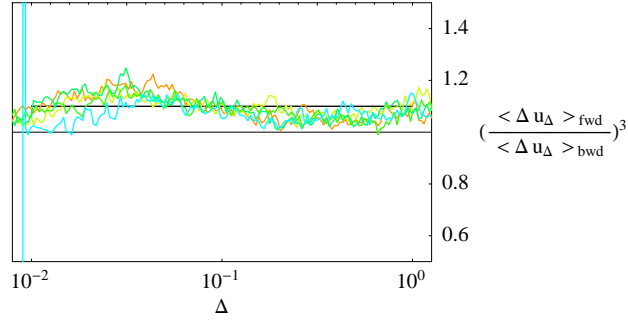}
\caption{Ratio of Richardson's constants $g_f/g_b$ from \eref{pairsep_diffusivity_ratio_fit_cond_final} for initial separations from the DNS data with $\Delta_0 \leq \eta_f$ within the inertial range ($-2 \approx \log\eta_f < \log\Delta < \log\L \approx -0.6$). The line colors are the same as in \fref{pairsep_dns_dtmssq_ratio}. In the absence of color note that the ratio is $>1$. The solid black line shows the best fit constant $g_f/g_b = 1.10$.}
\label{fig:pairsep_dns_UDeltaSep_ratio_0-4}
\end{figure}

\begin{table}[h!tbp]
\begin{center}
\begin{tabular}{cccc}
\hline\hline\hlinevspace
$\Delta_0$	& $g_f/g_b$  \\ \hline \hlinevspace
$\eta_f/16$	& $1.120 \pm 0.006$ \\
$\eta_f/8$	& $1.109 \pm 0.004$ \\
$\eta_f/4$	& $1.106 \pm 0.003$ \\
$\eta_f/2$	& $1.121 \pm 0.005$ \\
$\eta_f$	& $1.069 \pm 0.005$ \\ \hline \hlinevspace
combined	& $1.105 \pm 0.022$ \\
\hline\hline
\end{tabular} 
\end{center}
\caption{Best fit values of ratios of Richardson's constants for forward and backward DNS data and confidence intervals as found by \texttt{gnuplot 4.0} for separations within the inertial range ($-2 \approx \log\eta_f < \log\Delta < \log\L \approx -0.6$). Data with $\Delta_0 > \eta_f$ did not have good enough statistics within the inertial range to be considered for this fit. The uncertainty given for the individual fits is the asymptotic standard error given by gnuplot and the uncertainty given for the combined (averaged) values is the standard deviation $\sigma_{n-1}$ of the distribution of values above plus the average error above.}
\label{tab:pairsep_dns_fit_richardson_const_ratio}
\end{table}

The results from fitting \eref{pairsep_diffusivity_ratio_fit_cond_final} to the data are compatible with the previous value of $g_f/g_b$ (see \tref{pairsep_dns_fit_richardson_const_ratio}). It can also be seen from \fref{pairsep_dns_UDeltaSep_ratio_0-4} that the ratio $g_f/g_b$ is larger than unity for all five initial separations shown. We conclude that the ratio of Richardson's constants for this 2D DNS of isotropic homogeneous turbulence with resolution $N=3072$ is the weighted average of the values given in \tref{pairsep_dns_fit_richardson_const} and \tref{pairsep_dns_fit_richardson_const_ratio}: 
\begin{equation}
\label{eq:pairsep_richardson_const_ratio_final}
 \frac{g_f}{g_b} = (1.09 \pm 0.03) \, .
\end{equation}

\section{Discussion}

The main limitation of the present model has been shown repeatedly in this article. It currently models an infinite inertial range with a multi-scale stagnation point topology of infinitely wide scale range. However, in order for the model to accurately describe the systems available in experiment and DNS, it would be necessary to include finite range effects.

Some suggestions are to 
incorporate the separation behavior similar to Brownian motion for scales larger than the integral scale (\sref{pairsep_suitability_on_analy_sol}) and to include dissipative effects for separations that are smaller than the distance between stagnations points.

All results obtained from comparison of DNS with the model need to be discussed in this light. Thus, the value obtained by GV04 of $\alpha_\mathrm{GV}=1.3$ might turn out to be in agreement with Richardson's prediction of $\alpha_\mathrm{GV}=1.5$, when one takes into account that the range of scales in the multi-scale stagnation point topology is finite (\sref{pairsep_model_comparison}).


These finite range effects are also the reason that Richardson's $t^3$ law is not clearly present in the DNS with a resolution of $N=3072$. However, this does not mean that one cannot interpret the DNS data at all. In \sref{pairsep_richardson_const_conditioned}, we were able to use a novel scale-dependent approach which confirms Richardson's scalings \cite{Richardson:1926} in our finite range DNS even though the usual diagnostics and statistics do not show them clearly. 
Richardson's constants for the forward and backward case of this 2D DNS of isotropic homogeneous turbulence with resolution $N=3072$ are found to be
\begin{eqnarray}
 G_{\Delta,\mathrm{fwd}} &=& (1.066\pm 0.020) \, , \\
 G_{\Delta,\mathrm{bwd}} &=& (0.999\pm 0.007) \, .
\end{eqnarray}
These values are approximately 4 times lower than the Richardson constant $g \approx 3.8$ for a similar 2D DNS with similar resolution as obtained by Boffetta and Sokolov \cite{Boffetta:2002} using exit time statistics and approximately twice as large as the value of $g \approx 0.5$ obtained from fitting the $\propto t^3$ law to a 2D turbulence experiment by Jullien \etal~\cite{Jullien:1999}. The spread in obtained Richardson constants may be due to non-universality stemming from  different characteristics of the investigated flows, but might also stem from the varying approaches in obtaining them. The notion of a non-universal Richardson constant is plausible and supported by 
\begin{equation}
 G_\Delta \propto C_B^3\, C_s^{3/d} \, ,
\end{equation}
which can be derived from equations \eref{pairsep_B_const}, \eref{pairsep_tau_t} and \eref{pairsep_msqsep_analytical}. This predicts that the Richardson constant depends on the stagnation point number $C_s$ and the constant $C_B$, both of which are not necessarily universal.

Finally, the investigation of the two dimensional DNS in \sref{pairsep_time_asymm_2d} has shown qualitatively that the forward-/backward time asymmetry of turbulent pair separation is opposite in 2D to what it is in 3D. Furthermore, utilizing conditioning on scales rather than on time yields a sound quantitative value for the ratio of Richardson's constants in the present 2D turbulence DNS (\sref{pairsep_richardson_const_conditioned}):
\begin{equation}
 \frac{g_f}{g_b} = (1.09 \pm 0.03) \, .
\end{equation}
This can be compared to a value for 3D turbulence obtained experimentally in a previous study \cite{Berg:2006}:
\begin{eqnarray}
 \mathrm{present~2D~DNS}: & \qquad & \frac{g_b}{g_f} = (0.92 \pm 0.03) \\
 \mathrm{3D~experiment~\cite{Berg:2006}}: & \qquad & \frac{g_b}{g_f} = (2.1 \pm 0.3)
\end{eqnarray}

Berg \etal~\cite{Berg:2006} explained the time asymmetry in terms of the positive sign of the mean second eigenvalue of the strain tensor in 3D turbulence. Their argument would imply that $g_f/g_b = 1$ in incompressible 2D turbulence, which we have shown not to be the case. However, this does not imply that their argument is wrong. It implies that their mechanism cannot be the only one contributing to the observed time asymmetry of pair separation. We shall attempt to give a qualitative explanation for what might be a different and perhaps additional mechanism.



\begin{figure}[h!tbp]
\centering
\includegraphics{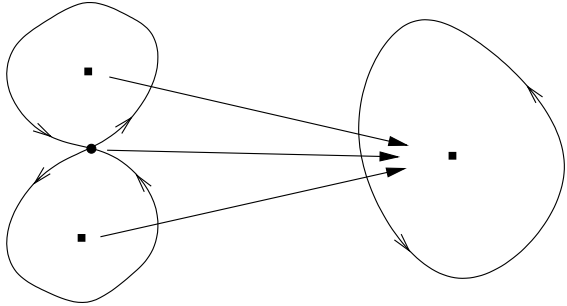}
\caption{\label{fig:pairsep_eddy_merge}
Diagram illustrating the merging of eddies. Shown on the left are two neighboring eddies with their elliptic ({\tiny $\blacksquare$}) and their hyperbolic ($\bullet$) stagnation points and a characteristic streamline. The right shows the larger eddy resulting from this merging process and its elliptic stagnation point. The merging involves the transformation of two elliptic and one hyperbolic to one elliptic stagnation point.}
\end{figure}

The mechanism causing this time asymmetry in two dimensional turbulence could have its origin in the asymmetry which exists between forward and inverse energy cascade. The latter is related to merging and thereby growing eddies in 2D turbulence, whereas the former is related to eddies breaking up into smaller eddies. 

It is known that in two dimensional turbulence, small vortices within each others proximity can merge into a vortex of a larger scale \cite{Davidson}. 
This idea of merging vortices has been explored for many decades \cite{Ayrton:1919,Fujiwhara:1923}. Picking up the notion of a particle pair's patron from \sref{pairsep_origmodel}, let us point out that it might be possible for a particle pair to move to a larger patron without encountering a hyperbolic stagnation point as presented in an earlier section. This alternative merging process would involve the annihilation of the small vortices and therefore the disappearance of their hyperbolic and elliptic zero-acceleration points and the creation of an elliptic zero-acceleration point for the emerging larger scale vortex, shown in \fref{pairsep_eddy_merge}. This dynamic process of a particle pair moving to a larger patron cannot be explicitly included in our model which is based on particle pairs encountering hyperbolic stagnation points.
However, it is possible to include the effect of this process in an effective value of either of the constants $C_s$, $C_B$ or both, which will be smaller if pairs separate whilst eddies break up into smaller eddies (as in our time-reversed 2D turbulence). Conversely, if pairs separate whilst eddies merge into larger eddies (as in forward time inverse-cascading 2D turbulence - see \fref{pairsep_eddy_merge}), the value of either $C_s$ and/or $C_B$ would be larger.

Summarizing, apart from the burst-like processes of separation and convergence described in \sref{pairsep_origmodel}, there might be an additional mechanism of merging vortices that also contributes to a particle pair's separation in 2D turbulence. This additional route to larger separation would strengthen the separation process in the natural time direction (forward), and would be reversed and therefore weakening the separation in the backwards case. Note that this process would be happening on time scales larger than the life time of the involved stagnation points. Note also that this process has not (yet) been directly observed in the context of pair separation and thus, is merely a suggested notion which consistently ties together all observations.

The inverse energy cascade of two dimensional turbulence is understood to describe the same phenomenon \cite{Davidson} of smaller vortices merging into larger ones. While nothing more than a handwaving argument, transferring this notion to three dimensional turbulence, where the energy cascade is associated with larger vortices breaking up into smaller ones, the separation process is depleted in the natural progression of time and strengthend in the backward case, giving a consistent qualitative picture for the observed time asymmetry in 2D and 3D turbulence.

\section{Conclusions}

An improved model of particle pair separation in flows with a multi scale stagnation point topology was devised, thus setting its predecessor \cite{GV04} on a sound mathematical foundation. The new model is able to model the observed time asymmetry provided that the effects of vortex-merging (\fref{pairsep_eddy_merge}) can be taken into account in effective values of $C_s$ and/or $C_B$. Furthermore, it has been argued that correction terms of higher order derivatives are necessary in the PDF evolution equation \eref{pairsep_PDF_PDE_ks} when considering separation moments of third order or higher.

The limitations and possible improvements to the present model are discussed, and the Richardson constant for the isotropic homogeneous 2D DNS turbulence is found to be of the same order of magnitude as previous comparable values \cite{Boffetta:2002,Jullien:1999}.

The time asymmetry is found to have opposing effects in two and three dimensional turbulence. It is suggested that the stronger forward separation in 2D turbulence might be caused by merging of eddies. Furthermore, the observed asymmetry is consistent with the assumption that the direction of the energy cascade in 2D and 3D turbulence is directly correlated with the direction of the time asymmetry.

\section{Acknowledgements}

The inspiration for this work was sparked by an issue raised at Arkady Tsinober's lecture series on \textit{Fundamental and Conceptual Aspects of Turbulent Flows} which he gave at Imperial College London, in January to March 2007. This research was supported by the Institute for Mathematical Sciences.

\bibliographystyle{plain}
\bibliography{../thesis/phd_references}

\begin{thebibliography}{10}

\bibitem{Ayrton:1919}
Hertha Ayrton.
\newblock On a new method of driving off poisonous gases.
\newblock {\em Proc. R. Soc. A}, 96:249, 1919.

\bibitem{Batchelor:1950}
G.~K. {Batchelor}.
\newblock {The application of the similarity theory of turbulence to
  atmospheric diffusion}.
\newblock {\em Quarterly Journal of the Royal Meteorological Society},
  76:133--146, April 1950.

\bibitem{Berg:2006}
Jacob Berg, Beat L\"{u}thi, Jakob Mann, and S\o{}ren Ott.
\newblock Backwards and forwards relative dispersion in turbulent flow: An
  experimental investigation.
\newblock {\em Physical Review E (Statistical, Nonlinear, and Soft Matter
  Physics)}, 74(1):016304, 2006.

\bibitem{Boffetta:2002}
G~Boffetta and I~M Sokolov.
\newblock Statistics of two-particle dispersion in two-dimensional turbulence.
\newblock {\em Physics of Fluids}, 14(9):3224, 2002.

\bibitem{BurdenFaires}
Richard~L Burden and J~Douglas Faires.
\newblock {\em Numerical Analysis}.
\newblock Brooks Cole, 7th revised edition, 2001.

\bibitem{Corrsin:1952}
Stanley Corrsin.
\newblock Heat transfer in isotropic turbulence.
\newblock {\em Journal of Applied Physics}, 23(1):113, January 1952.

\bibitem{Davidson}
P.~A. Davidson.
\newblock {\em Turbulence, an introduction for scientists and engineers}.
\newblock Oxford University Press, 2004.

\bibitem{Davila:2003}
J~D\'{a}vila and J~C Vassilicos.
\newblock Richardson's pair diffusion and the stagnation point structure of
  turbulence.
\newblock {\em Physical Review Letters}, 91(14):144501, October 2003.

\bibitem{Durbin:1980}
P~A Durbin.
\newblock A stochastic model of two-particle dispersion and concentration
  fluctuations in homogeneous turbulence.
\newblock {\em Journal of Fluid Mechanics}, 100(2):279, 1980.

\bibitem{Flohr:2000}
P~Flohr and J~C Vassilicos.
\newblock A scalar subgrid model with flow structure for large-eddy simulations
  of scalar variances.
\newblock {\em Journal of Fluid Mechanics}, 407:315, 2000.

\bibitem{Frisch:1999}
U~Frisch, A~Mazzino, A~Noullez, and M~Vergassola.
\newblock Lagrangian method for multiple correlations in passive scalar
  advection.
\newblock {\em Physics of Fluids}, 11(8):2178, August 1999.

\bibitem{Fujiwhara:1923}
S~Fujiwhara.
\newblock On the growth and decay of vortical systems.
\newblock {\em Q.J.R. Met. Soc.}, 49(206):75, Apr 1923.

\bibitem{Fung:1998}
J~C~H Fung and J~C Vassilicos.
\newblock Two-particle dispersion in turbulentlike flows.
\newblock {\em Physical Review E (Statistical, Nonlinear, and Soft Matter
  Physics)}, 57(2):1677, 1998.

\bibitem{GV04}
Susumu Goto and J~C Vassilicos.
\newblock Particle pair diffusion and persistent streamline topology in
  two-dimensional turbulence.
\newblock {\em New Journal of Physics}, 6:65, 2004.

\bibitem{Jimenez:1993}
Javier Jim\'{e}nez, Alan~A Wray, Philip~G Saffman, and Robert~S Rogallo.
\newblock The structure of intense vorticity in isotropic turbulence.
\newblock {\em Journal of Fluid Mechanics}, 255:65, 1993.

\bibitem{Jullien:1999}
Marie-Caroline Jullien, J\'er\^ome Paret, and Patrick Tabeling.
\newblock Richardson pair dispersion in two-dimensional turbulence.
\newblock {\em Physical Review Letters}, 82(14):2872--2875, Apr 1999.

\bibitem{Luethi:2007}
Beat L\"{u}thi, Jacob Berg, S\o{}ren Ott, and Jakob Mann.
\newblock Self-similar two-particle separation model.
\newblock {\em Physics of Fluids}, 19:045110, 2007.

\bibitem{Nicolleau:2004}
F~Nicolleau and G~Yu.
\newblock Two-particle diffusion and locality assumption.
\newblock {\em Physics of Fluids}, 16(7):2309, 2004.

\bibitem{Ott:2000}
S\o{}ren Ott and Jakob Mann.
\newblock An experimental investigation of the relative diffusion of particle
  pairs in three-dimensional turbulent flow.
\newblock {\em Journal of Fluid Mechanics}, 422:207, 2000.

\bibitem{Richardson:1926}
L.~F. Richardson.
\newblock Atmospheric diffusion shown on a distance-neighbour graph.
\newblock {\em Proc. R. Soc. A}, 110:709, 1926.

\bibitem{Sawford:2005}
Brian~L. Sawford, P.~K. Yeung, and Michael~S. Borgas.
\newblock Comparison of backwards and forwards relative dispersion in
  turbulence.
\newblock {\em Physics of Fluids}, 17(9):095109, 2005.

\bibitem{She:1991}
Zhen-Su She, Eric Jackson, and Steven~A. Orszag.
\newblock Structure and dynamics of homogeneous turbulence: Models and
  simulations.
\newblock {\em Proceedings: Mathematical and Physical Sciences},
  434(1890):101--124, 1991.

\bibitem{Thomson:2003}
David~J Thomson.
\newblock Dispersion of particle pairs and decay of scalar fields in isotropic
  turbulence.
\newblock {\em Physics of Fluids}, 15(3):801, March 2003.

\bibitem{Yeung:2002}
P~K Yeung.
\newblock Lagrangian investigations of turbulence.
\newblock {\em Annual Review of Fluid Mechanics}, 34:115, January 2002.

\end{thebibliography}

\end{document}